\documentclass{aastex62}
\usepackage{gensymb}
\usepackage{amssymb}

\received{January 26, 2019}
\revised{February 24, 2019}
\accepted{March 4, 2019}
\submitjournal{AJ}

\shorttitle{RAMSES II}
\shortauthors{Angeloni et al.}

\begin{document}

\title{RAMSES II -- RAMan Search for Extragalactic Symbiotic Stars\\Project concept, commissioning, and early results from the science verification phase}

\correspondingauthor{Denise~R.~Gon\c{c}alves (Project PI)}
\email{denise@astro.ufrj.br}

\author[0000-0001-7978-7077]{Rodolfo~Angeloni}
\affil{Instituto de Investigaci\'on Multidisciplinar en Ciencia y Tecnolog\'ia, Universidad de La Serena, Av. ​R. Bitr\'an 1305, La Serena, Chile}
\affiliation{Departamento de F\'isica y Astronom\'ia, Universidad de La Serena, Av. J. Cisternas 1200, La Serena, Chile}

\author{Denise~R.~Gon\c{c}alves}
\affiliation{Observat\'orio do Valongo, Universidade Federal do Rio de Janeiro, Ladeira Pedro Antonio 43, 20080-090, Rio de Janeiro, Brazil}

\author{Stavros~Akras}
\affiliation{Observat\'orio do Valongo, Universidade Federal do Rio de Janeiro, Ladeira Pedro Antonio 43, 20080-090, Rio de Janeiro, Brazil}

\author{German~Gimeno}
\affiliation{Gemini Observatory, Southern Operations Center, Casilla 603, La Serena, Chile}

\author{Ruben~Diaz}
\affiliation{Gemini Observatory, Southern Operations Center, Casilla 603, La Serena, Chile}

\author{Julia~Scharw{\"a}chter}
\affiliation{Gemini Observatory, Northern Operations Center, 670 N. A'ohoku Place, Hilo, HI 96720, USA}

\author{Natalia~E.~Nu\~{n}ez}
\affiliation{Instituto de Ciencias Astron\'{o}micas, de la Tierra y del Espacio (ICATE-CONICET), Av. Espa\~{n}a Sur 1512, J5402DSP, San Juan, Argentina}

\author[0000-0002-2647-4373]{Gerardo~Juan~M.~Luna}
\affiliation{CONICET-Universidad de Buenos Aires, Instituto de Astronom\'{\i}a y F\'{\i}sica del Espacio (IAFE), Av. Inte. G\"{u}iraldes 2620, C1428ZAA, Buenos Aires, Argentina}
\affiliation{Universidad de Buenos Aires, Facultad de Ciencias Exactas y Naturales, Buenos Aires, Argentina}
\affiliation{Universidad Nacional Arturo Jauretche, Av. Calchaqu\'{\i} 6200, F. Varela, Buenos Aires, Argentina}

\author{Hee-Won~Lee}
\affiliation{Department of Physics and Astronomy, Sejong University, Seoul 05006, Republic of Korea}

\author{Jeong-Eun~Heo}
\affiliation{Department of Physics and Astronomy, Sejong University, Seoul 05006, Republic of Korea}

\author[0000-0003-4827-9402]{Adrian~B.~Lucy}
\affiliation{Columbia University, Dept. of Astronomy, 550 West 120th Street, New York, NY 10027, USA}
\affiliation{LSSTC Data Science Fellow}

\author{Marcelo~Jaque~Arancibia}
\affiliation{Departamento de F\'isica y Astronom\'ia, Universidad de La Serena, Av. J. Cisternas 1200, La Serena, Chile}

\author{Cristian~Moreno}
\affiliation{Gemini Observatory, Southern Operations Center, Casilla 603, La Serena, Chile}

\author{Emmanuel~Chirre}
\affiliation{Gemini Observatory, Southern Operations Center, Casilla 603, La Serena, Chile}

\author{Stephen~J.~Goodsell}
\affiliation{Gemini Observatory, Northern Operations Center, 670 N. A'ohoku Place, Hilo, HI 96720, USA}
\affiliation{Department of Physics, Durham University, South Road, Durham, DH1 3LE, UK}

\author[0000-0002-3704-3368]{Piera~Soto King}
\affiliation{Departamento de F\'isica y Astronom\'ia, Universidad de La Serena, Av. J. Cisternas 1200, La Serena, Chile}

\author{J.~L.~Sokoloski}
\affiliation{Columbia University, Dept. of Astronomy, 550 West 120th Street, New York, NY 10027, USA}
\affiliation{Large Synoptic Survey Telescope Corporation, 933 North Cherry Ave, Tucson, AZ 85721, USA}

\author{Bo-Eun~Choi}
\affiliation{Department of Physics and Astronomy, Sejong University, Seoul 05006, Republic of Korea}

\author{Mateus~Dias~Ribeiro}
\affiliation{Observat\'orio do Valongo, Universidade Federal do Rio de Janeiro, Ladeira Pedro Antonio 43, 20080-090, Rio de Janeiro, Brazil}



\begin{abstract}
Symbiotic stars (SySts) are long-period interacting binaries composed of a hot compact star, an evolved giant star, and a tangled network of gas and dust nebulae. They represent unique laboratories for studying a variety of important astrophysical problems, and have also been proposed as possible progenitors of SNIa. 
Presently, we know 257 SySts in the Milky Way and 69 in external galaxies. However, these numbers are still in striking contrast with the predicted population of SySts in our Galaxy. 
Because of other astrophysical sources that mimic SySt colors, no photometric diagnostic tool has so far demonstrated the power to unambiguously identify a SySt, thus making the recourse to costly spectroscopic follow-up still inescapable. In this paper we present the concept, commissioning, and science verification phases, as well as the first scientific results, of RAMSES II -- a Gemini Observatory Instrument Upgrade Project that has provided each GMOS instrument at both Gemini telescopes with a set of narrow-band filters centered on the Raman OVI 6830 \AA\ band. Continuum-subtracted images using these new filters clearly revealed known SySts with a range of Raman OVI line strengths, even in crowded fields. RAMSES II observations also produced the first detection of Raman OVI emission from the SySt LMC 1 and confirmed Hen 3-1768 as a new SySt -- the first photometric confirmation of a SySt. Via Raman OVI narrow-band imaging, RAMSES II provides the astronomical community with the first purely photometric tool for hunting SySts in the local Universe.
\end{abstract}


\section{Introduction} \label{sec:intro}
Symbiotic stars (hereafter SySts) are long-period interacting binaries composed of a hot compact star -- generally but not necessarily a white dwarf (WD) -- and an evolved giant star, whose mutual interaction via accretion processes is at the origin of the extended emission recorded from radio to X-rays.
Nowadays, SySts represent unique laboratories for studying a variety of important astrophysical problems and their reciprocal influence: e.g., nova-like thermonuclear outbursts (Skopal 2015), formation and collimation of jets (Angeloni et al. 2011; Tomov 2003), PNe morphology (Corradi 2003), and variable X-ray emission (Luna et al. 2013), among others.
As binary systems, they offer a powerful benchmark to study the effect of binary evolution on the nucleosynthesis, mixing, and dust mineralogy
that characterize the giant companion, likely different from what expected in single RGB and AGB stars (Marigo et al. 2008; Marigo \& Girardi 2007). Importantly, they are among the most promising candidates as progenitors of SNIa (e.g., I{\l}kiewicz et al. 2018c; Meng \& Han, 2016; Dimitriadis et al. 2014; Dilday et al. 2012).\\

The most up-to-date SySt catalog (Akras et al. 2019a) lists 257 objects in the Milky Way and 66 in external galaxies: it is larger by almost a factor of two with respect to the previous compilation by Belczy{\'n}ski et al. (2000), which almost twenty-years ago included a total of 188 confirmed SySts. However, the growing number of observed SySts is still in striking contrast with the predicted population expected in our Galaxy. According to different theoretical estimates SySts may number between $\sim$10$^3$ (Lu et al., 2012; Allen 1984) and a few 10$^5$ (Magrini et al. 2003). For example, Magrini et al. (2003) suggest that the expected number of SySts in a given galaxy would be comparable to $\sim$0.5\% of the total number of its RGB and AGB populations (Table~\ref{tab:predicted}).

One of the reasons for the discrepancy in the number of observed vs. expected SySts also stems from the fact that, historically, this class of variable stars has been defined on the basis of purely spectroscopic criteria (I{\l}kiewicz \& Miko{\l}ajewska 2017; Belczyński et al. 2000). Because many other stellar sources appear to mimic SySt colors (PNe, Be and T Tauri stars, CVs, Mira LPVs, etc. -- see, e.g., Figs. 1 \& 2 in Corradi et al. 2008; Akras et al. 2019b), no photometric diagnostic tool has so far demonstrated the power to unambiguously identify a SySt, thus making the recourse to costly spectroscopic follow-up still inescapable.\\

In recent years, several research groups around the globe have both started extensive observing surveys aimed at discovering and characterizing SySts in external galaxies -- particularly in the Magellanic Clouds (I{\l}kiewicz et al. 2018a) -- and, at the same time, have explored new approaches (Lucy et al. 2018) and techniques (e.g., machine-learning algorithms -- Akras et al. 2019b) to optimize the classification criteria for distinguishing SySts from their astrophysical ``impostors''. Nonetheless, in all cases, a confirmation spectrum is still compulsory to obtain a trustworthy identification of a new member of the symbiotic family selected from the (ever growing) lists of potential candidates. And so far, there remains a significant ``waste of spectrographic time spent on mimics'' (I{\l}kiewicz et al. 2018a).\\

\begin{deluxetable*}{cccccc}
\tablecaption{Predicted vs. spectroscopically confirmed number of extragalactic SySts in a sample of LG galaxies\label{tab:predicted}}
\tablewidth{0pt}
\tablehead{
\colhead{Galaxy} & \colhead{Distance$^a$} &\colhead{M$^b$}         & \colhead{Predicted$^c$} & \colhead{Observed}\\
\colhead{}    & [kpc]      &\colhead{[M$_\odot$]} & \colhead{\# of SySts} &\colhead{\# of SySts}
}
\startdata
NGC 147    & 730$\pm$101    & 5.5$\times$10$^7$     & 2\,800& -   \\
NGC 185    & 616$\pm$26    & 6.6$\times$10$^8$      & 4\,200& 1$^d$    \\
NGC 205    & 824$\pm$27    & 7.5$\times$10$^8$       & 17\,000& 1 $^e$   \\
M 31    & 792$\pm$440    & 2-4.0$\times$10$^{11}$    & 660\,000& 31 $^f$   \\
M 32    & 771$\pm$63    & 1.1$\times$10$^9$       & 19\,000&  -  \\
M 33    & 883$\pm$246    &0.8-1.4$\times$10$^{10}$   & 45\,000& 12$^g$    \\
Fornax    & 138$\pm$5    & 6.8$\times$10$^7$       & 500    &  -  \\
Leo I    & 254$\pm$17    & $>$2.0$\times$10$^7$    & 200&   - \\
Leo II    & 233$\pm$15    & 1.1$\times$10$^7$       & 50&  -  \\
Draco    & 76$\pm$6    & 1.7$\times$10$^7$     & 10& 1$^h$    \\
\enddata
\tablecomments{$a$: from the NASA/IPAC Extragalactic Database. $b$: from Mateo (1998). $c$: taken from Magrini et al. (2003). 
$d$: from Gon\c{c}alves et al. (2012). $e$: from Gon\c{c}alves et al. (2015a). 
$f$: from Miko{\l}ajewska et al. (2014). $g$:  from Miko{\l}ajewska et al. (2017). $h$: from Munari (1991).}
\end{deluxetable*}

Raman spectroscopy offers an invaluable diagnostic tool to constrain the accretion processes and geometry in SySts (Lee et al. 2016; Heo et al. 2016; Seker{\'a}{\v s} \& Skopal 2015; Shore et al. 2010). The two intense Raman OVI bands at $\lambda\lambda$6830, 7088 \AA\AA\ are so unique to the symbiotic phenomenon that their presence has been commonly used as a sufficient criterion for classifying a star as symbiotic, even in those cases where the cool companion appears to be hiding.  From an observing point of view, whenever present, the $\lambda$6830 \AA\ band appears as a rather strong feature: it is among the 10 most intense lines in the optical, able to reach up to 5\% of the intensity of H$\alpha$ (Allen 1980; Schmid 1989; Akras et al. 2019a), and it is easily recognizable because of broad (FWHM$\approx$20 \AA) and rather composite profiles (double or even triple-peaks are usually seen in high-resolution spectra - Heo et al. 2016). Despite the uncertain detection of Raman-scattered OVI bands in a handful of possibly non-symbiotic objects -- such as very young PNe (Sahai \& Patel 2015), one B[e] star (Torres et al. 2012), and the classical CO nova V339 Del (Shore et al. 2014; Skopal et al. 2014) -- their presence is still a virtually clear-cut proof of a \textit{bona fide} SySt. 

These unique spectroscopic features are due to Raman-scattering of the O VI $\lambda\lambda$1032, 1038 \AA\AA\ resonance doublet by neutral H (Schmid 1989). Given the high ionization potential of O$^{5+}$ (114 eV), Raman-scattered OVI lines indicate the presence of a strong ionizing source, i.e., of a very hot WD. High temperatures can be achieved if the accreted material is burned as it is accreted onto the WD surface. These SySts are known as \textit{shell-burning} symbiotics (e.g. Luna et al. 2013), and for them the WD temperature is a function of its mass. It is therefore understandable that 100\% of the hottest shell burning SySts, detected as super-soft X-ray sources ($\alpha$-types), display Raman-scattered OVI bands in their optical spectra, while all the unambiguously non-burning sources ($\delta$-types) do not (Luna et al. 2013; Akras et al. 2019a). 

Interestingly, the presence of such a hot and luminous WD implies a tight relation between the HeII 4686 line and the Raman OVI 6830 band. From the overall sample of spectroscopically confirmed SySts, it is inferred that whenever the Raman OVI line is present, the HeII 4686 line is also present (Akras et al. 2019a). The simultaneous detection of these two lines in a stellar object would therefore provide an unquestionable identification of a SySt.\\ 

Raman features alone are a sufficient but not necessary condition to classify an object as symbiotic. Allen (1980) already noted a general tendency that Akras (2019a) has just confirmed: about 55\% of known SySts in the Milky Way show Raman-scattered OVI bands. For the other galaxies, the presence of Raman emission is at the moment confirmed in 92\% of the SySt sample in the Small Magellanic Cloud, 57\% in the Large Magellanic Cloud (LMC), 42\% in M33 and 52\% in M31. Moreover, even if the numbers are still too low to support any statistical argument, there are a handful of other Local Group galaxies in which Raman-emitter SySts have already been discovered  (Table~\ref{tab:predicted}).  Raman-scattered OVI bands appear therefore a very suitable tool to discover shell-burning SySts in the Milky Way and Local Group galaxies.\\ 

In this paper we present the technical concept (Sect.~\ref{sec:design}), commissioning (Sect.~\ref{sec:characterization}), and science verification (SV) phase with its very first scientific results (Sect.~\ref{sec:sv}), of RAMSES II -- a Gemini Observatory Instrument Upgrade Project that has provided each Gemini Multi-Object Spectrograph (GMOS -- Hook et al. 2004; Gimeno et al. 2016; Scharw{\"a}chter et al. 2018) at both Gemini telescopes with a set of narrow-band filters centered on the Raman OVI 6830 \AA\ band and an adjacent portion of the local continuum. 
It aims at discovering and characterizing the symbiotic population of the Milky Way and Local Group galaxies via Raman OVI narrow-band imaging, providing the astronomical community with the very first tool entirely based on purely photometric criteria for hunting SySts in the local Universe. A general discussion emphasizing the novelty and power of this ambitious project appears in Section~\ref{sec:disc}, while concluding remarks follow in Section~\ref{sec:remarks}.

\section{Filter design} \label{sec:design}
Given the general astrophysical context presented in the Introduction, 
and convinced that the idea of searching for unidentified SySts through narrow-band Raman OVI emission was worth a proper feasibility study, we faced a three-fold issue: first, identify the astronomical facility most suitable to accomplish the science goals; then, transform the general inputs from the science case into specific technical requirements for the filter design; and eventually (and probably most importantly), locate the funding channel able to support the project -- which in the meantime was given the name of \textit{RAMan Search for Extragalactic Symbiotic Stars}: RAMSES II.
The first and third points were jointly solved at the end of 2016 thanks to a Gemini Observatory Instrument Upgrade Program (Sect.~\ref{sec:iup}), which also immediately put constraints on the filter design (Sect.~\ref{sec:techspec}).

\subsection{Gemini Instrument Upgrade Program} \label{sec:iup}
In the constant effort of upgrading its existing operational instruments to keep them scientifically competitive and to create new instrument capabilities, the Gemini Observatory announced in 2015 the first call of its Instrument Upgrade Program\footnote{http://www.gemini.edu/sciops/future-instrumentation-amp-current-development/instrument-upgrade-projects} (IUP -- Diaz et al. 2018, 2016), a funding source for community-created, science-driven proposals.
Gemini's baseline plan is to provide for one small project ($\sim$100,000 USD) every year and one medium project ($\sim$500,000 USD) every other year. Every selected project is awarded up to one night (10 hours) of observing time to be used to test and demonstrate the scientific potential of the upgraded instrument.

The RAMSES II project was awarded after the 2016 IUP call (under MOU \#20173, signed on behalf of the proponent team by the PI D.~R.~Gon\c{c}alves, Observat\'orio do Valongo, Universidade Federal do Rio de Janeiro, Brazil), and it proposed the design and manufacturing of one set of narrow-band Raman OVI filters for each GMOS at the two Gemini telescopes.

\subsection{Filter requirements} \label{sec:techspec}
The requirements submitted to the vendor (Asahi Spectra USA Inc.) were the result of the combined interplay between the top-level technical specifications imposed by the GMOS instruments\footnote{https://www.gemini.edu/sciops/instruments/gmos/imaging/filters/user-supplied-filters}, and the specific science case, which provided the most suitable  central wavelength $\lambda_c$ and FWHM for both the on- and off-band filters (Table~\ref{tab:specifications}). \\

\begin{table}[!hb]
\begin{center}
\caption{GMOS filter requirements}
\label{tab:specifications}
\begin{tabular}{ccc}
\hline
\hline
Technical specifications &  RAMSES II values\\
\hline
Central wavelength $\lambda_c$ & OVI filter: 6835 \AA \\
FWHM & 50 \AA \\
Substrate & Fused silica or similar \\
Operating temperature range &  -10$\degree$ to 20$\degree$ C \\
Substrate diameter &	160 +0.0 mm -0.2mm \\
Substrate thickness &	10.0 mm \\
Mechanical thickness &	11.0 mm\\
Coated aperture &	$>$150 mm \\
Flatness &	$^a\lambda$/4 (p-v) at 633 nm over any 100 mm $\varnothing$ patch\\
Parallelism &	$<$30 arcseconds \\
Cosmetic quality &	$^a$60-40 scratch dig, no visible pinholes \\
Angle of incidence & $^b$4.12$\degree$ \\
\hline
\end{tabular}
\tablecomments{The OVIC filter specifications are identical to the OVI ones listed above, apart from a different $\lambda_c$ of 6780 \AA. \\\textit{a}: Feature meets or exceeds these specifications. \\
\textit{b}: There are two angles of incidence used in GMOS: an original specification of 3 degrees, and a change in 2009 to 4.12 degrees. RAMSES II filters utilize the 4.12-degree angle of incidence as it significantly reduces ghosting.}
\end{center}
\end{table}

The central wavelength $\lambda_c$ = 6835 \AA\ of the on-band filters (hereafter OVI) was selected on the basis of the mean position of the Raman OVI band as observed in a large set of galactic and extragalactic SySts. The central wavelength $\lambda_c$ = 6780 \AA\ of the off-band filters (hereafter OVIC) was chosen by carefully inspecting the very diverse morphology of SySt continua in the spectral region between the [SII] 6717/6731 doublet and the HeI 7065 line, as reported in the literature: particularly useful has been the multi-epoch spectrophotometric atlas by Munari \& Zwitter (2002, hereafter MZ02 -- see also Fig.~\ref{fig:mzprofile}), who published optical spectra for 130 galactic and extragalactic SySts. The presence of the telluric O$_2$ B band at $\lambda \approx$ 6870 \AA\ (Groppi et al. 1996) was our main reason for centering the OVIC filters at $\lambda_c$=6780 \AA\, i.e., on the blue side of the OVI ones.
The filter FWHM (50 \AA) was finally set on the basis of the typical width of the Raman band profiles ($\approx$20 \AA, Schmid 1989) and of the observed, intrinsic dispersion of central wavelengths due to local kinematic effects peculiar to any SySt.

\section{Commissioning Phase \label{sec:characterization}}
In this section we summarize the different characterization tests executed during the RAMSES II early commissioning phase, following the strategy highlighted in the Acceptance Test Plan (v2.2) and detailed in the Acceptance Test Report.

\subsection{Optical lab characterization\label{subsec:lab}}
The two filter sets -- 2$\times$(OVI, OVIC) -- were shipped by the vendor to the Gemini Observatory Southern Operations Center in La Serena, Chile. Upon their arrival in February 2018, the filters were visually inspected for relevant physical defects, their physical diameters and thickness were carefully determined, and their optical transmission was finally measured with a CARY 500 spectrophotometer at the Gemini optical lab.

\begin{figure}[ht!]
\plotone{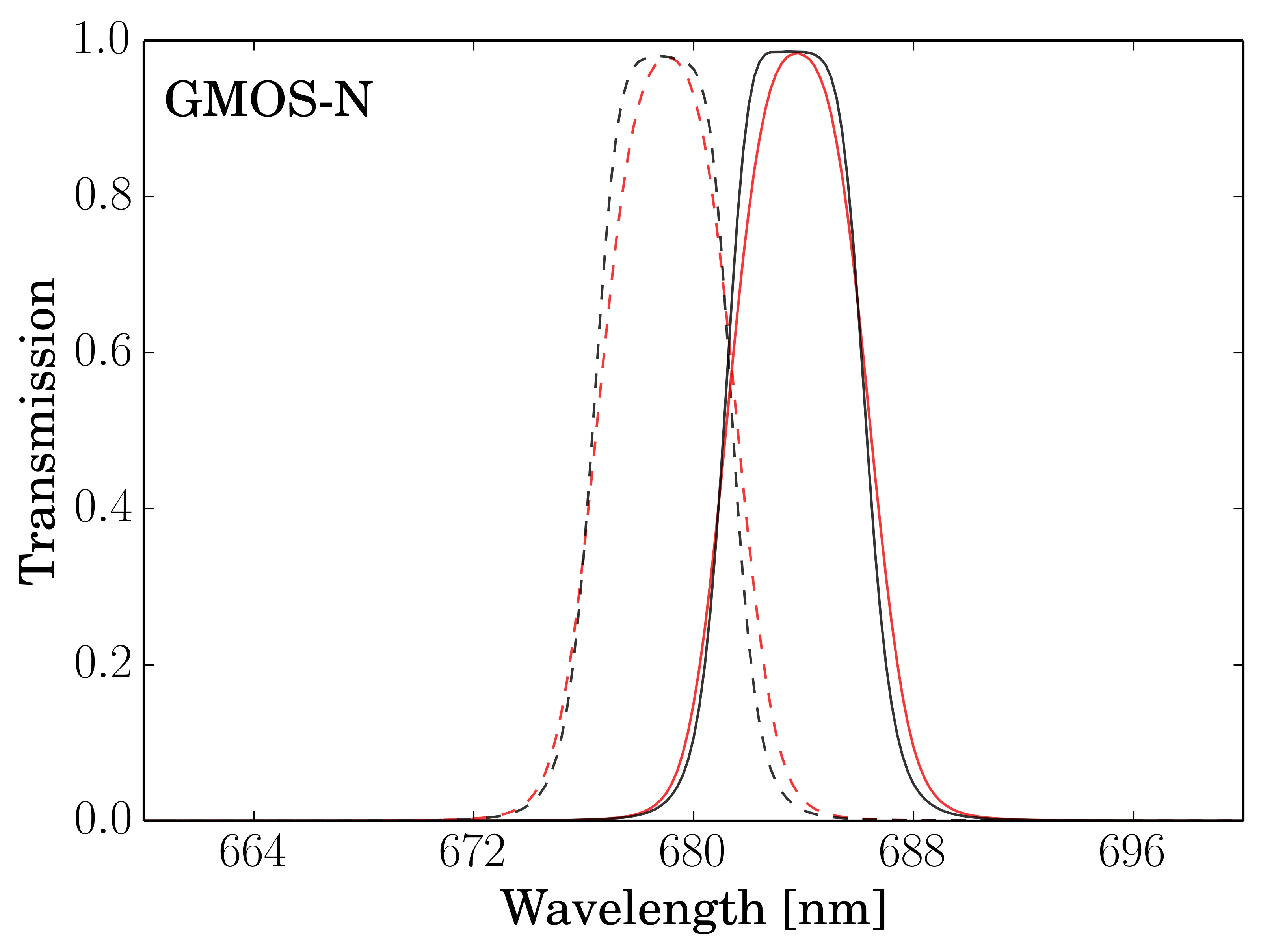}
\caption{Transmission curves of the RAMSES II filters installed in GMOS-N. The solid lines refer to the OVI filter, while the dashed lines refer to the OVIC filter; the black curves are the transmission as provided by the manufacturer, while the red curve is the transmission measured in the Gemini optical lab during the early commissioning phase. The transmission curves of the GMOS-S RAMSES II filters look very similar, and are not shown here (but see Table~\ref{tab:lab}). \label{fig:trans}}
\end{figure}

Both filter sets show very similar transmission curves (Fig.~\ref{fig:trans}): the filters match or even exceed the required specifications in terms of cosmetics, physical properties and optical properties (Table~\ref{tab:lab}). In particular, it is worth reporting that the total transmission is $>$90\%, a value that exceeds the original requirement and is better than the transmission of the current GMOS H$\alpha$ filters.

	\begin{table}[!hbt]
		\begin{center}
		\caption{Measured central wavelength $\lambda_c$ and FWHM of RAMSES II filters}
		\label{tab:lab}
		\begin{tabular}{ccc}
			\hline
			\hline
			Filter &  $\lambda_c$ & FWHM\\
			ID & [\AA] & [\AA] \\
			\hline
			GN OVI 6835 & 6838.0$\pm$0.1 & 49.0$\pm$0.2 \\
			GN OVIC 6780 & 6790.3$\pm$0.1 & 48.2$\pm$0.2 \\
			GS OVI 6835 & 6840.4$\pm$0.1 & 49.0$\pm$0.2 \\
			GS OVIC 6780 & 6784.6$\pm$0.1 & 48.1$\pm$0.2 \\
			\hline
		\end{tabular}
		\end{center}
	\end{table}

\subsection{Day-time tests \label{subsec:daytests}}
After the optical lab characterization, one filter set was installed in GMOS-S (March 2018), and the other one was shipped from Chile to Hawaii and installed in GMOS-N (May 2018), for both starting the respective day-time tests with the Gemini CALibration unit (GCAL). 
The next commissioning step was to calibrate the filter surface focus offset in the instrument and to check the image quality with the final calibration. The image quality was measured by illuminating the focal plane array with GCAL through a pinhole grid mask.
The images were bias subtracted and then the FWHM and the radius of an aperture that encompassed 85\% of the total encircled energy (EE85) for the pin-hole sources were measured with the \texttt{gemseeing} task within the Gemini IRAF \texttt{gemtools} package. The main aim was to verify that the OVI and OVIC filters produce image quality with an EE85 diameter no more than 20\% worse than that measured on contemporaneous H$\alpha$ images. In all cases, the image quality (FWHM and EE85) were proven fully within the requirements.


\subsection{Observing strategy, data reduction and analysis} \label{subsec:dra}
The on-sky data required to proceed with the commissioning (Sect.~\ref{subsec:onsky}) and SV (Sect.~\ref{sec:sv}) phases were taken with GMOS-S (GMOS-N) through the engineering program GS-2018A-ENG-156 (GN-2018A-ENG-52) executed during March (June/July) 2018. 
At Gemini-South, we extended the early SV thanks to the GMOS B4  poor weather program GS-2018A-Q-405 (long-slit mode, 1 arcsec slit, R400 grating centered at 6800 \AA, executed in April-May, 2018), which allowed us to obtain timely spectroscopic follow-up of some puzzling sources (Sections~\ref{subsubsec:smp},~\ref{subsubsec:v366 car},~\ref{subsubsec:lmc1}), and to image in Raman OVI a few recently announced SySt candidates on which to further test the RAMSES II performance (Sections~\ref{subsubsec:hen3},~\ref{subsec:cand}). \\

All data were taken through the standard Gemini software (\texttt{Observing Tool} and \texttt{Seqexec}). In the OT, we implemented the same observing strategy for both the above mentioned phases. For each OVI filter, we adopted an \textit{n}-step random dither pattern (usually, \textit{n=3} or \textit{n=4}), that was then identically replicated for the OVIC filter. The entire (OVI+OVIC) observing sequence was taken within the same scheduling block, in order to minimize any possible seeing (i.e., PSF) variation between the images. This strategy has guaranteed an easy and very reliable continuum-subtraction (OVI-OVIC) without the need of implementing more sophisticated and time consuming differential imaging techniques (see Section~\ref{sec:disc} for more details).\\

The raw data were processed using the GMOS workflow available in the \textit{Image Reduction and Analysis Facility} (\texttt{IRAF}) \textit{Gemini package} (v1.14), which takes care of the most common reduction steps including bias subtraction, flat fielding and mosaicking. 
For both filters, we thus obtained the corresponding reduced (i.e., mosaicked and combined) frames that were in turn astrometrically registered in order to execute the last reduction step, i.e., the subtraction of the OVIC continuum image from the OVI on-band one, after multiplication by an empirically determined scaling factor (usually very close to 1).\\

Following the filter characterization process, for those SySts which showed a Raman OVI 6830 detection in both the RAMSES II images and the B4 GMOS-S spectra (see Sects.~\ref{subsubsec:lmcs147}, \ref{subsubsec:lmc1}, \ref{subsubsec:hen3}), we compared the corresponding band equivalent width $|W_\lambda|$. From the long-slit spectra, we directly measured the equivalent width with the task \texttt{splot} available from the \texttt{noao.onedspec} package in \texttt{IRAF}. In the RAMSES II images, we applied the definition of equivalent width as $|W_\lambda|=\int (1-\frac{F_\lambda}{F_0}) d\lambda$: in this (necessarily approximated) case, $F_{\lambda}$ is the counts of the target PSF in the OVI filter, $F_{0}$ is the counts of the target PSF in the OVIC filter (that would therefore represent the band ``underlying'' continuum) and $d\lambda$ is simply taken to be the filter FWHM (50 \AA). \\

Finally, it may be worth clarifying that some artifacts -- like hot pixels and columns, cosmic rays, residuals from very saturated stars, charge smearing effects -- from which the instrument detectors suffered in a few cases were not always perfectly removed and may still appear in the final images (see Figure~\ref{fig:standard} for an illustrative example): they only affect the image overall aesthetics and, needless to say, are not related to the Raman filters, nor modify our results, but demonstrate once more the robustness of the method.

\subsection{On-sky tests} \label{subsec:onsky}
At this stage, RAMSES II was ready for the ``first light'' on-sky. Selected sparse and crowded sky regions were observed under photometric and good seeing conditions (i.e., CC50\%-ile, IQ70\%-ile in Gemini's jargon) with the OVI, OVIC and H$\alpha$ filters: the latter filter was used as a comparison reference baseline. 
We characterized once more the image quality over the entire GMOS FoV, this time on-sky, by measuring the FWHM and EE85 for the source star PSFs; evaluated the sky background count rates; determined the system relative throughput and the filter preliminary zero points (the latter reported in Table~\ref{tab:sens}); constrained the exact size of the unvignetted FoV; and verified that no significant ghosting is present when pointing towards very bright sources.

The results of these first on-sky tests confirmed that the filters are fully compliant with the originally specified technical requirements and provide imaging data comparable in quality to the existing GMOS H$\alpha$ filters.\\

\begin{figure}[ht!]
\plottwo{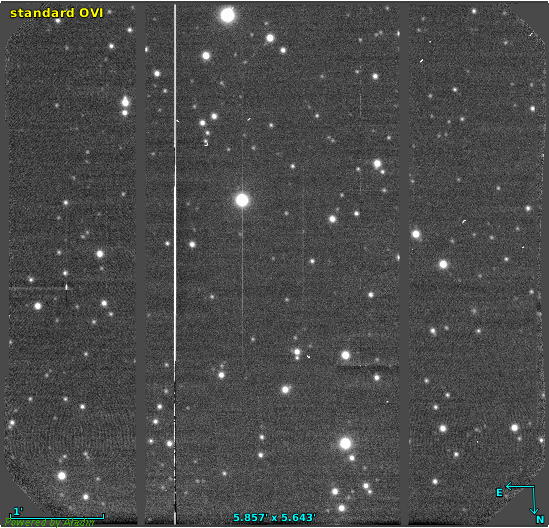}{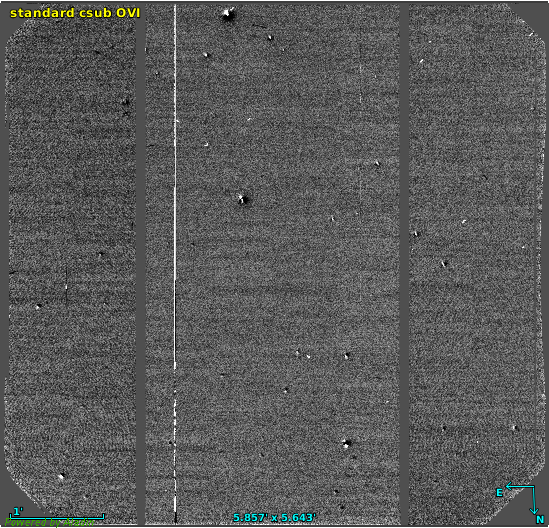}
\caption{First light of RAMSES II, obtained with GMOS-S on March 14, 2018: OVI (left panel) and continuum-subtracted (i.e., OVI-OVIC; right panel) frames of the sky field around the photometric standard star TYC 9054-1091-1.\label{fig:standard}}
\end{figure}

Figure~\ref{fig:standard} shows the first light of RAMSES II, obtained with GMOS-S on March 14, 2018: it is the sky field around the photometric standard star TYC 9054-1091-1. The left panel is the OVI frame, while the right panel is the continuum-subtracted (OVI-OVIC) one. Since the two filters are very similar and the sky conditions between the two consecutive exposures (both of t$_{exp}$=5 sec) remained virtually identical, the PSFs in the two images are in turn virtually identical: as to-be-expected, the continuum-subtracted frame does not show any significant signal.

\begin{table}[!ht]
		\begin{center}
		\caption{Preliminary zero point values of RAMSES II filters}
		\label{tab:sens}
		\begin{tabular}{ccc}
			\hline
			\hline
			Filter & Zero Point\\
			ID &   [mag] \\
			\hline
			GN OVI 6835  & 23.8 \\
			GN OVIC 6780 & 23.8\\
			GS OVI 6835  & 23.1 \\
			GS OVIC 6780 & 23.1 \\
			\hline
		\end{tabular}
		\tablecomments{The ZPs were obtained by averaging the instrumental magnitudes of many stars over a wide range of counts: considering the different spectral classes of the objects, we estimated a ZP uncertainty of $\sim$5\%.}
		\end{center}
	\end{table}

\section{Science Verification Phase} \label{sec:sv}

\subsection{Target selection} \label{subsec:svts}
\begin{figure}[ht!]
\plotone{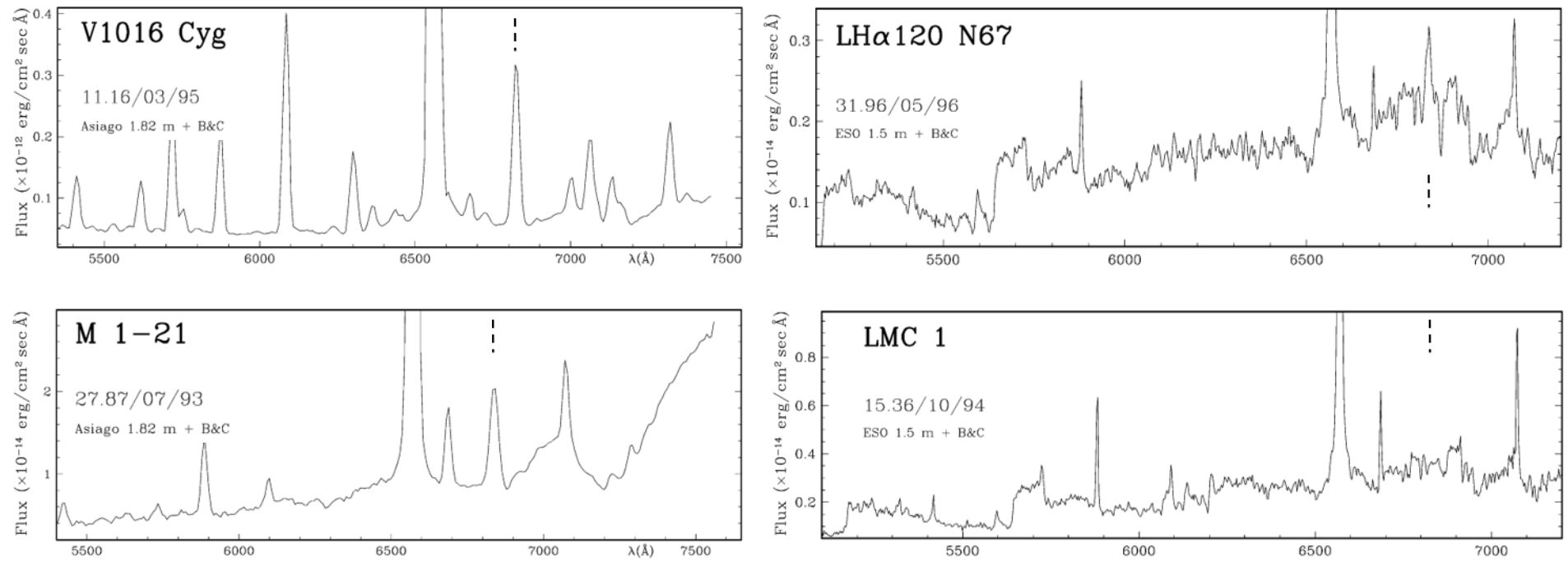}
\caption{Flux-calibrated optical spectra of a representative sample of SySts (adapted from MZ02). The dashed lines mark the position of the OVI Raman 6830 band. The presence of a telluric absorption due to the O$_2$ B band at $\lambda\approx$ 6870 \AA\ (clearly visible in the spectrum of LH$\alpha$ 120 N67) was the main reason for centering the OVIC filter at $\lambda_c$=6780 \AA. \label{fig:mzprofile}}
\end{figure}

The criteria adopted for the target selection during the subsequent SV phase took into account several factors. First of all, we wanted to extensively explore the different spectral types encountered in SySts, from the extreme of those systems in which the M giant companion is so absorbed by its own dust shell as not to be directly visible in the optical (e.g., V1016 Cyg, Section~\ref{subsubsec:v1016}), down to the ``yellow'' SySt in which the donor star is of quite earlier spectral type (i.e., F to K -- as is the case of SMP LMC 88, Section~\ref{subsubsec:smp}). We wanted also to test the ability of our method to recover Raman OVI emission of different strengths and on top of local continua of very different (nebular plus stellar) shape (Fig.~\ref{fig:mzprofile}). And of course, we had to consider both the individual target visibility and the total available time in the observing window assigned by Gemini Observatory for the AT and SV phases. In order to monitor the filter performance under possible cases of false positives, we also included two well-characterized SySts known not to show Raman emission (i.e., CM Aql and LMC 1). A very helpful visual guide was offered by the MZ02 spectrophotometric atlas, complemented with the Belczy{\'n}ski et al. (2000) and Akras et al. (2019a) catalogs. It is worth noting that the MZ02 atlas, which also guided the conceptual design of RAMSES II in the early stages of the project (Sect.~\ref{sec:design}), presents data taken back in the 1990's: for a non-negligible fraction of SySts no more recent spectra are available, and therefore any effects of (spectroscopic) variability remain virtually unknown, including the ill-constrained variability of their Raman emission. \\

Figure~\ref{fig:mzprofile} shows four representative types of spectral energy distribution usually encountered in SySts, and in which the photospheric signatures of the cool giant become more and more dominant. In the top left panel, V1016 Cyg is a Raman-emitting SySt in which the Mira (M7, Muerset \& Schmid 1999, hereafter MS99) is absorbed in the optical and the continuum appears relatively flat. 
At the top right, LH$\alpha$ 120 N67 is a carbon SySt belonging to the LMC (Muerset et al. 1996, hereafter Mu96). The carbon-rich nature of the cool component is evident from the spectrum, with the presence of both the Raman OVI 6830 band and the O$_2$ telluric band at $\lambda \approx$ 6870 \AA: as explained in Section~\ref{sec:techspec}, the presence of such absorption was our main reason for centering the OVIC filters on the blue side of the OVI band.
At the bottom left, M1-21 is another example of a Raman-emitting SySt whose cool component has been classified to be of spectral type M6 (MS99).
Finally, in the bottom right panel appears LMC 1, a symbiotic star in LMC. Classified as another carbon-rich star, it is an example of a SySt in which no Raman emission has ever been recorded.\\

In the end, a total of 19 objects, representative of the very diverse phenomenology in which Raman emission appears in SySts, were observed between March and July 2018.  In the following, we present a selected sample of SySts observed during the SV phase at both Gemini telescopes. It is meant to exemplify the filters' performance when targeting different spectral types and Raman band relative strengths (Sections~\ref{subsec:er} and~\ref{subsec:newsr}), and to illustrate the kind of spurious detection we may face when using the Raman OVI filters alone (Section~\ref{subsec:cand}). The journal of observations for the (confirmed and candidate SySt) targets discussed below appears in Table~\ref{tab:targets}.

\begin{deluxetable*}{cccccccc}
\tablecaption{List of program stars and journal of observations \label{tab:targets}}
\tablewidth{0pt}
\tablehead{
\colhead{Name$^a$} & \colhead{$\alpha_{J2000}$} &\colhead{$\delta_{J2000}$}  & \colhead{Obs. date} & \colhead{$t_{exp}$} & \colhead{Fig. \#}& \colhead{Spectral}\\
\colhead{}    & $h \; m \; s$      &\colhead{$\degree \; ' \; ''$} & \colhead{2018/mm/dd} &\colhead{\#$\times$[sec]} & This study & Ref.$^b$
}
\startdata
LHA120 S154 & 04 51 50.469  & -75 03 35.36 & 03/15 & 3$\times$120 & \ref{fig:lmcs154}& I19\\
LHA120 S147 & 04 54 03.473 & -70 59 32.18 & 04/02 & 4$\times$90 & \ref{fig:lmcs147}& Mu96\\
LMC 1 & 05 25 01.106 & -62 28 48.78 & 03/14 & 3$\times$60 &  \ref{fig:lmc1},\ref{fig:lmc1spec} & MZ02\\
LHA120 N67 & 05 36 07.576 &  -64 43 21.34 & 03/14 & 3$\times$30 &\ref{fig:lmcn67}& MZ02 \\
SMP LMC 88 & 05 42 33.193 & -70 29 24.08 & 03/14 & 3$\times$120 & \ref{fig:smp},\ref{fig:smpspec}& I18b\\
Sanduleak's star & 05 45 19.569 & -71 16 06.72 & 03/15 & 3$\times$60&\ref{fig:sandy}& H16 \\
\textit{ASASSN-V J081823.00-111138.9} & 08 18 23.001 & -11 11 38.95 & 05/10 &3$\times$60 & - & -\\
V366 Car & 09 54 43.284 & -57 18 52.40 & 03/14 & 3$\times$30 & \ref{fig:v366} & MZ02\\
\textit{CD-28 10578} & 14 18 28.908 & -28 39 03.73 & 05/13 & 4$\times$25& - & -\\
\textit{NSVS J1444107-074451} & 14 44 10.676	& -07 44 49.42 & 05/13 & 4$\times$60 & - & -\\
\textit{GSC 09276-00130}  & 17 18 09.290 & -67 57 26.00 & 05/14 & 4$\times$60 & - & -\\
M 1-21   & 17 34 17.218 &   -19 09 22.81  & 07/09 & 3$\times$10  & \ref{fig:m121} & MZ02 \\
V1016 Cyg & 19 57 05.019 & +39 49 36.09  & 06/23 & 3$\times$5 & \ref{fig:v1016} & MZ02 \\
\textit{Hen 3-1768}   & 19 59 48.418 &  -82 52 37.49  & 05/14 &  4$\times$30 & \ref{fig:hen3}& L18
\enddata
\tablecomments{$^a$Candidate SySts appear in \textit{italic}; $^b$(I19) I{\l}kiewicz et al. 2019; (Mu96) Muerset et al. 1996; (MZ02) Munari \& Zwitter 2002; (I18b) I{\l}kiewicz et al. 2018b; (H16) Heo et al. 2016; (L18) Lucy et al. 2018.}
\end{deluxetable*}

\subsection{Early results from previously known Raman OVI emitters} \label{subsec:er}

\subsubsection{V1016 Cyg} \label{subsubsec:v1016}
V1016 Cyg is one of the most studied SySt. More than a hundred papers in the SAO/NASA ADS include its name in the title, offering a panchromatic picture of a system that since its nova-like outburst in 1964 has not ceased to capture the attention of the professional astronomical community and, in the latest years, of the ever growing population of experienced amateur astronomers\footnote{http://www.astrosurf.com/aras/Aras\_DataBase/Symbiotics/V1016Cyg.htm}.
To give even a short review of this complex astrophysical system is beyond the scope of this work: the interested reader will find in the astronomical literature many excellent reviews, some offering comprehensive observational summaries like the paper by Arkhipova et al. (2016) that celebrates half a century from the 1964 outburst. 

V1016 Cyg is an outstanding natural laboratory of Raman scattering processes in astrophysics. Apart from the intense Raman OVI bands, it is one of the very few objects (along with RR Tel) in which other Raman-scattered lines have been detected: e.g., HeII 6545 (Lee et al. 2003), HeII 4850 (Jung \& Lee 2004), HeII 4332 (Lee 2012), NeVII 4881 (Lee et al. 2014).
Being such a well-characterized SySt, it was compulsory to include it in our northern target list.\\

\begin{figure}[ht!]
\plottwo{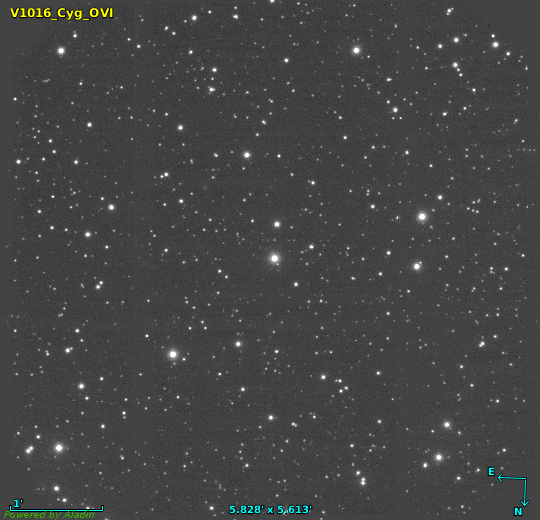}{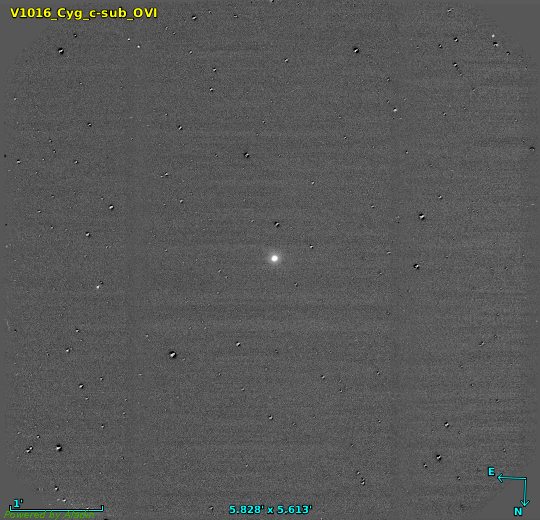}
\caption{RAMSES II images of V1016 Cyg, obtained with GMOS-N on June 23, 2018. Left panel: Raman OVI frame. Right panel: corresponding continuum-subtracted image. The target is located at the center of the field.\label{fig:v1016}}
\end{figure}

The RAMSES II glance at V1016 Cyg (just 3$\times$5 sec exposures in each OVI and OVIC filter, obtained on June 23, 2018) is presented in Fig.~\ref{fig:v1016}: the left panel shows the OVI frame, and the right panel displays the continuum-subtracted (OVI 6835 -- OVIC 6780) one. The target sits at the center of the $\sim$5.5 arcmin$^2$ GMOS-N field of view: despite the significant decrease in the Raman OVI 6830 flux registered between 1995 and 2013 (Arkhipova et al. 2015), its emission clearly stands out also in these first, very promising, RAMSES II snapshots.

\subsubsection{LH$\alpha$120 N67} \label{subsubsec:lmcn67}
\begin{figure}[ht!]
\plottwo{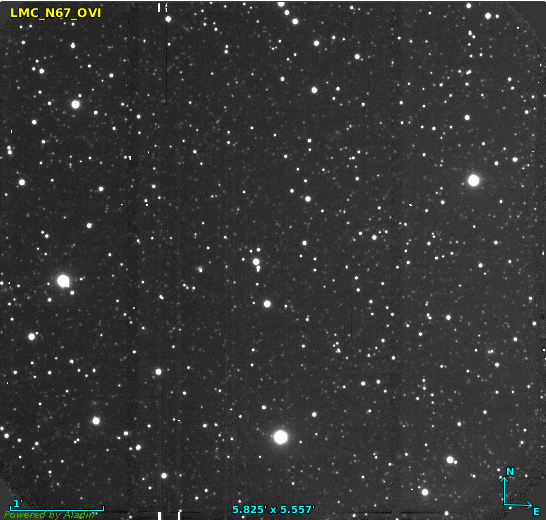}{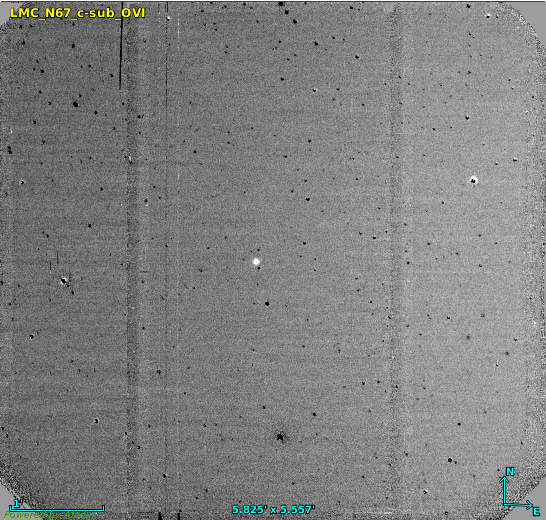}
\caption{As in Fig. \ref{fig:v1016}, but for LH$\alpha$120 N67. The Raman emission of this LMC SySt (observed with GMOS-S on March 14, 2018) was easily recovered.\label{fig:lmcn67}}
\end{figure}

As has been already mentioned in Section~\ref{subsec:svts}, LH$\alpha$120 N67 is a carbon-rich SySt in LMC. We are not aware of any dedicated studies of its Raman emission, but both spectra shown by MZ02 (and reproduced in our Fig.~\ref{fig:mzprofile}) and Mu96 (their Fig.~2) clearly reveal the presence of a fairly intense Raman OVI 6830 band. There were two main motivations for adding this object to our southern target list. We wanted to test the robustness of our method in recovering the Raman OVI 6830 emission i) on a far-from-smooth local continuum\footnote{This jagged continuum is unlike those of V1016 Cyg or Sanduleak's star.}, and ii) when in the proximity of a strong O$_2$ 6870 telluric feature. The RAMSES II images of N67 were taken on March 14, 2018 (Fig.~\ref{fig:lmcn67}): it was extremely reassuring to see surfacing its Raman emission in just 3 minutes of overall time on target.

\subsubsection{M 1-21} \label{subsubsec:hen2247}
The MZ02 spectrum of M 1-21 (\textit{aka} Hen 2-247) was shown in Fig.~\ref{fig:mzprofile}, and is that of a Raman-emitter SySt with an M6 giant as donor component. Since the spectral type distribution of SySts peaks between M5 and M6 according to MS99 (their Fig.~6), it is an educated guess to take M 1-21 as a first-order approximation of the kind of local continuum on which the Raman bands would likely appear. This was the main reason for including it in our target list for Gemini North.

\begin{figure}[ht!]
\plottwo{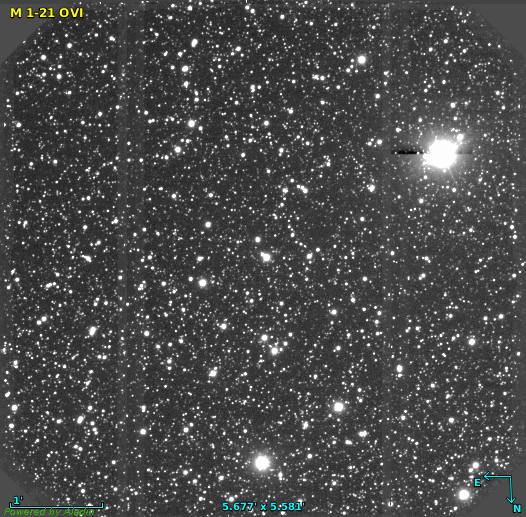}{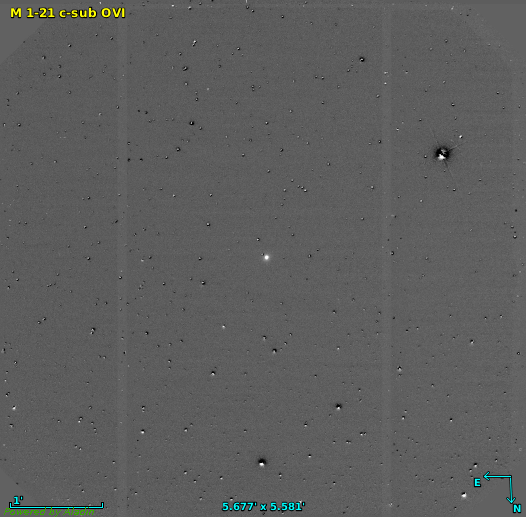}
\caption{As in Fig. \ref{fig:v1016}, but for M 1-21 (observed on July 9, 2018). 60 seconds of Gemini North telescope time were sufficient to isolate the Raman emission of this SySt from the the other stars visible in this relatively crowded field. \label{fig:m121}}
\end{figure}

Interestingly enough, M 1-21 is one of the very few SySts for which the orbital parameters are particularly well-constrained: the orbital period P$_{orb}$=898$\pm$5 days, as given by Fekel et al. (2008), was in fact derived by combining their own infrared radial velocities with spectropolarimetry of Raman-scattered OVI emission lines previously obtained by Harries \& Howarth (2000). Spectropolarimetry observations of Raman features are essential in providing two orbital elements that cannot be determined from radial velocities: the inclination, \textit{i}, and the position angle of the line of nodes, \textit{$\Omega$} (Schmid \& Schild 1994, 1990). M 1-21 was observed by RAMSES II with GMOS-N on July 09, 2018 (Fig.~\ref{fig:m121}): an overall time on target of just 60 sec (3$\times$10 sec per filter) was sufficient to also recover this Raman emitter.

\subsubsection{Sanduleak's star} \label{subsubsec:sandy}
Sanduleak's star is probably the most famous example of an astrophysical object whose symbiotic classification relies almost entirely on the presence of the Raman OVI bands. As a matter of fact, no clear signature of a late-type star is detectable in the optical-NIR spectra. 

The object, located in LMC, is extraordinary \textit{per se}, since it triggers the largest bipolar stellar jet known to date (Angeloni et al. 2011; Camps-Fari{\~n}a et al. 2018). Its photometric behavior is also rather peculiar: for more than two decades, Sanduleak's star has been monotonically fading at a rate of $\sim$0.03 mag/year in all optical bands, suggesting that it is probably still recovering from some (unnoticed) nova-like outburst (Angeloni et al. 2014). A detailed modeling of its strong Raman OVI emission-line profiles based on far-UV and optical high-resolution spectra has been presented by Heo et al. (2016).\\ 

\begin{figure}[ht!]
\plottwo{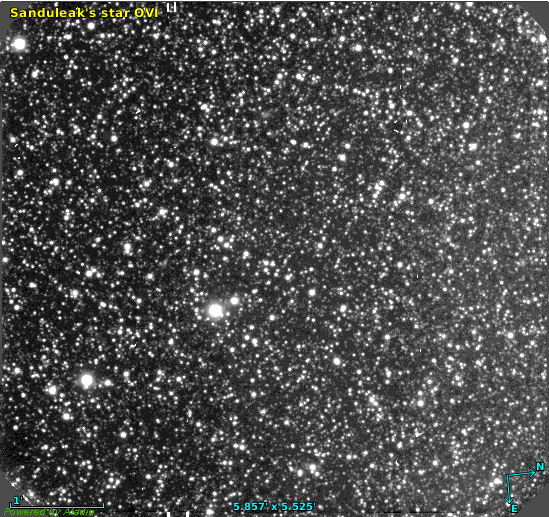}{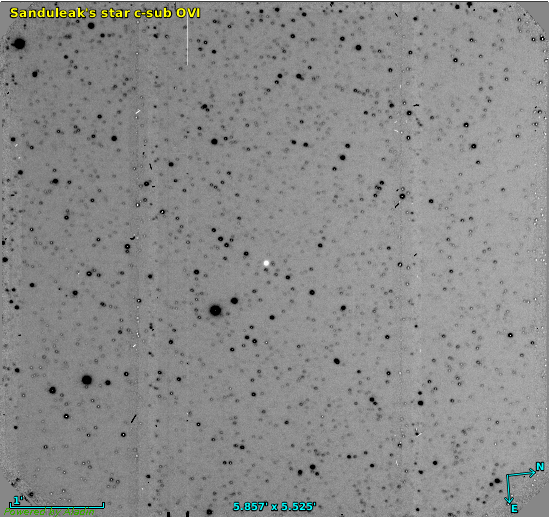}
\caption{As in Fig. \ref{fig:v1016}, but for Sanduleak's star (observed with GMOS-S on March 15, 2018). The image subtraction technique was able to easily recover the target also in a particularly crowded field like this one in LMC.\label{fig:sandy}}
\end{figure}

Because of its intrinsic fascination and (as for the other objects located in LMC) optimal visibility during the SV phase with GMOS-S, it was quite natural to include it in our southern target list. We observed it on March 15, 2018 with 3$\times$60 sec exposure time in each OVI and OVIC filter (for a total time on target of just 6 minutes). The on-band and the corresponding continuum-subtracted images are shown in Fig.~\ref{fig:sandy}: RAMSES II not only promptly recovered the object, but did it in a particularly crowded field, which points to the strong reliability of our new methodology.

\subsubsection{SMP LMC 88} \label{subsubsec:smp}
SMP LMC 88 is another example of a star whose symbiotic nature has been  unambiguously confirmed by the identification of its Raman OVI bands. Originally classified as a planetary nebula, I{\l}kiewicz et al. (2018b) have recently suggested that the object must actually be a ``yellow'' symbiotic, i.e., a SySt in which the cool component is a giant of earlier spectral type (K-type in this particular case) than in traditional SySts.

Its spectrum reveals a wealth of emission lines on a rather flat continuum, and photometric and spectroscopic variability is clearly present on time scales of just a few years. Interestingly enough, Table~1 and Fig.~4 of I{\l}kiewicz et al. (2018b) show that the intensity of Raman OVI 6830 (absent before 2013) has been constantly decreasing from $F$(OVI 6830)=4.7$\times$10$^{-15}$ erg cm$^{-2}$ s$^{-1}$ in January 2013 to $F$(OVI 6830)=6.8$\times$10$^{-16}$ erg cm$^{-2}$ s$^{-1}$ in October 2017, and that its ratio with adjacent emission lines (e.g., [SII] 7631, HeI 7065, [ArIII] 7135) has also decreased in like vein.\\

\begin{figure}[ht!]
\plottwo{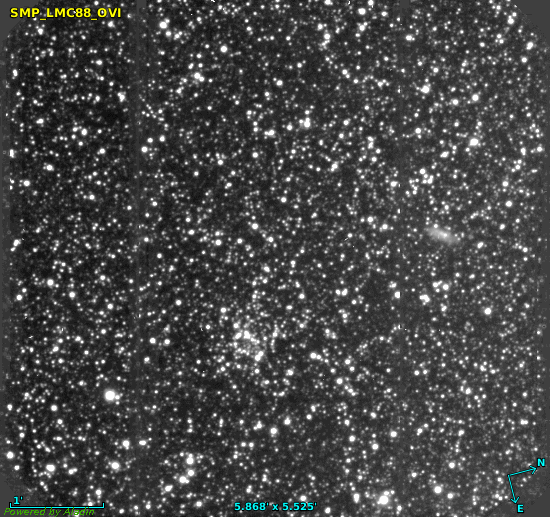}{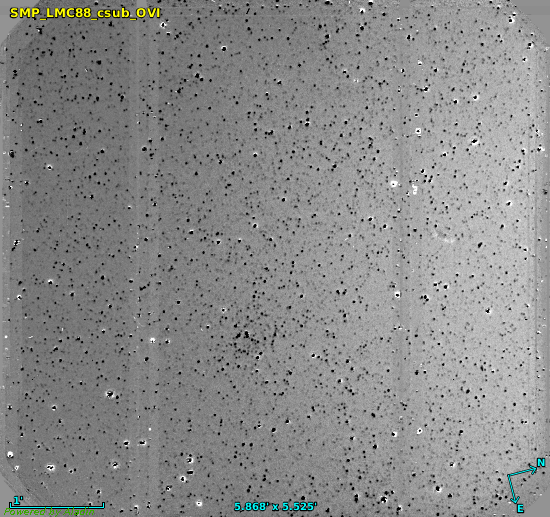}
\caption{As in Fig. \ref{fig:v1016}, but for SMP LMC88 (first observed on March 14, 2018). In this case, no Raman emission was detected. 
The few ``white spots'' in the continuum-subtracted image have very well-defined PSFs, suggesting that they could potentially constitute new sources of Raman OVI emission. They will be the subject of a forthcoming follow-up investigation. The residual in correspondence to the background galaxy in the mid-right side of the field is a reduction artifact, and not a positive detection (see text for details).\label{fig:smp}}
\end{figure}

\begin{figure}[ht!]
\plotone{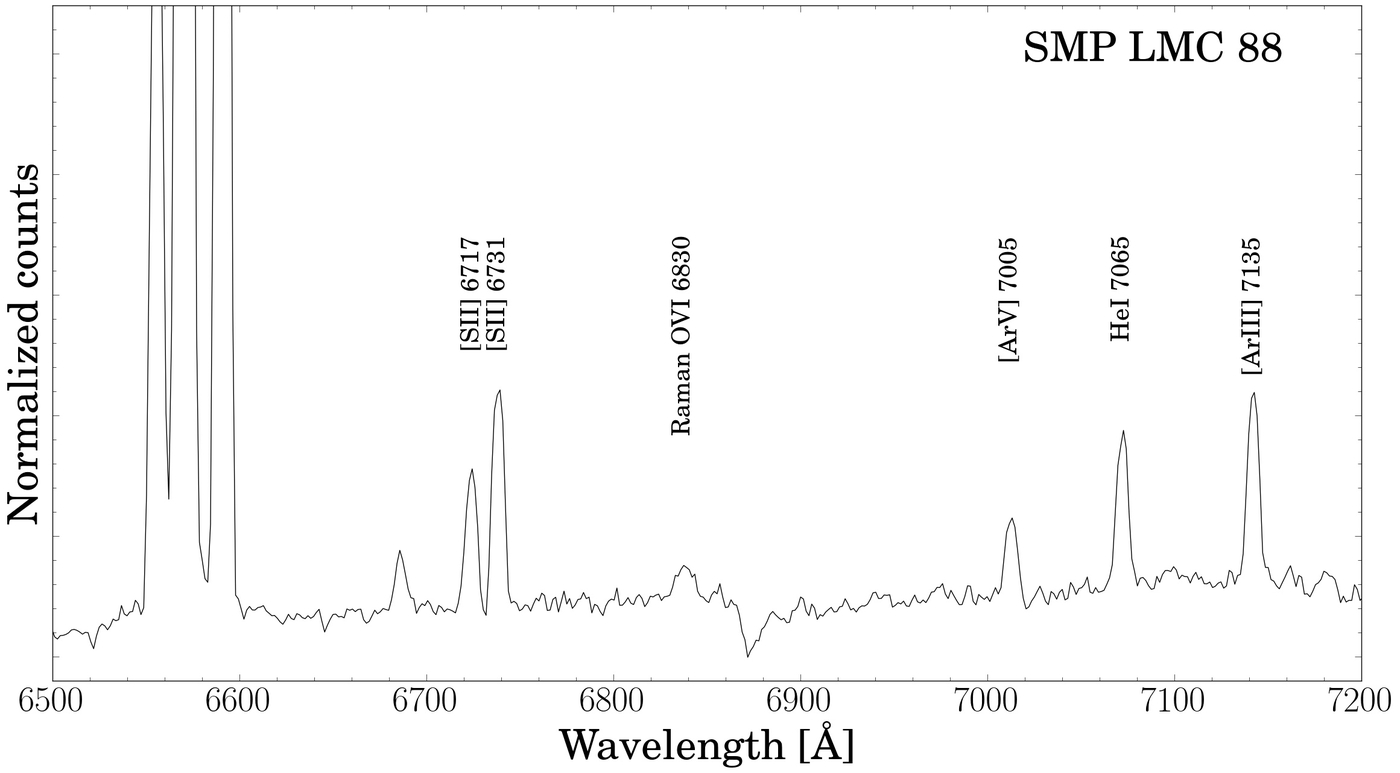}
\caption{GMOS-S optical spectrum of SMP LMC 88 obtained on April 02, 2018: the Raman OVI 6830 band is again very weak, while the OVI 7088 band is absent. It is instructive to compare this spectrum with the ones displayed in Fig.~4 of I{\l}kiewicz et al. (2018b). \label{fig:smpspec}}
\end{figure}

We first observed SMP LMC 88 on March 14, 2018 with RAMSES II. The 2$\times$120 sec exposures in both OVI and OVIC filters, presented in Fig.~\ref{fig:smp}, did not detect any Raman emission. Due to the strong spectral variability of the target, we opted for a spectroscopic follow-up, and observed it again on April 02, 2018 under B4 conditions: the GMOS spectrum, here displayed as Fig.~\ref{fig:smpspec}, confirmed that the Raman OVI 6830 ($\lambda_c$=6838 \AA, FWHM  $\approx$ 11 \AA, $|W_\lambda|\approx$ 3 \AA) was very weak, and the OVI 7088 band was absent. This non-detection is a first example of the potential application of our method in monitoring the time variability of Raman OVI emission. At the same time, it helped us to characterize RAMSES II detection limits.\\

Finally, it is worth noting that the white spots in the continuum-subtracted image of Fig.~\ref{fig:smp} have very well-defined PSFs, suggesting that they could potentially constitute new sources of Raman OVI emission. No additional narrow-band images nor spectra are currently available for these anonymous objects, which will be the subject of a forthcoming follow-up investigation. It is also worth noticing that the adopted PSF subtraction strategy may leave residuals in correspondence to very faint, spatially extended structures, as is the case for the background galaxy appearing towards the mid-right side of the field: these residuals are clearly reduction artifacts and not positive detections.

\subsubsection{LH$\alpha$120 S147} \label{subsubsec:lmcs147}
Another intriguing object in LMC is LH$\alpha$120 S147. Morgan \& Allen (1988), who first proposed it as a member of the symbiotic family, comment on the presence of relatively strong Raman OVI features (whose carrier at that time was still unidentified) by showing the entire optical spectrum of S147 in their Fig.~1. Mu96 also discuss the system in their seminal review of extragalactic SySts.\\

\begin{figure}[ht!]
\plottwo{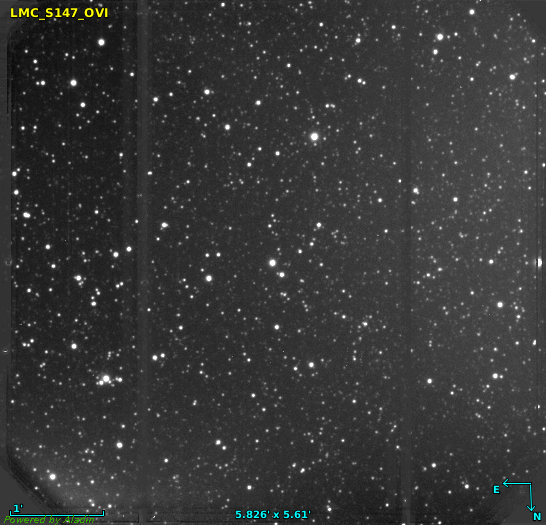}{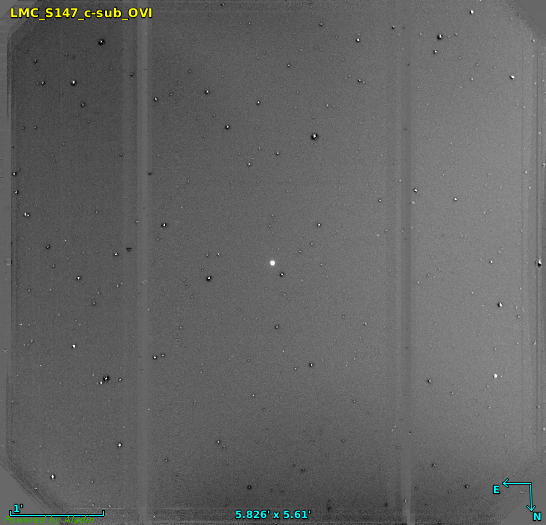}
\caption{As in Fig. \ref{fig:v1016}, but for LH$\alpha$120 S147 (observed with GMOS-S on April 02, 2018). The Raman source is clearly visible at the center of the field.\label{fig:lmcs147}}
\end{figure}

Figure~\ref{fig:lmcs147} shows the on-band and the continuum-subtracted images of S147 observed on the night of April 02, 2018 with a total integration time (per filter) of 6 minutes. Its Raman emission was promptly recovered and, following the method described at the end of Section~\ref{subsec:dra}, we obtained a $|W_\lambda^{Ram}|\approx$ 18 \AA. The follow-up B4 spectrum (executed on April 22, 2018 but not shown here) confirmed that the Raman OVI 6830 band was indeed still present and quite strong ($\lambda_c$=6837 \AA, FWHM  $\approx$ 16 \AA, $|W_\lambda|\approx$ 17 \AA). Assuming that it had not changed significantly in the 20 days between the RAMSES II images and the B4 spectrum, the slight difference in the two $|W_\lambda|$ values are likely due to the effect of a variable seeing that, on April 02, increased from 0.96 arcsec in the OVI images, to 1.06 arcsec in the OVIC images. The better seeing in the first image set is indeed consistent with a RAMSES II $|W_\lambda^{Ram}|$ larger than the $|W_\lambda|$ measured from the B4 spectrum.

\subsubsection{LH$\alpha$120 S154} \label{subsubsec:lmcs154}
A further example of Raman OVI variable emission is represented by LH$\alpha$120 S154. This object, presented in the  H$\alpha$-emission catalog of Henize (1956), was first studied in some detail by Remillard et al. (1992), who traced its fast evolution from a low-excitation ``Fe II star'' to a high-excitation state reminiscent of a SySt. Their Figure~3 shows, along with a dramatic increase of the excitation level of the emission lines from 1984 to 1989, a Raman OVI 6830 band so variable in strength that it is virtually absent in the February 1988 spectrum, then clearly visible only 10 months after, then almost gone again in Feb 1989.

Very recently, I{\l}kiewicz et al. (2019) have presented a detailed photometric and spectroscopic monitoring of this interesting but poorly studied object. Their Figure~1 shows a sequence of six optical spectra taken between 2005 and 2015, that confirm the strong spectral variability already emphasized by Remillard et al. (1992). The Raman OVI 6830 band is present in the 2005, 2006, and 2007 spectra (its strength increasing by more than 25\% in the first two years), but absent in the 2008 and 2009 spectra. In their most recent spectrum, taken on October 28, 2015, the Raman OVI 6830 band appears again, but at an intensity level that is just $\sim$70\% of the original 2005 value.\\

\begin{figure}[ht!]
\plottwo{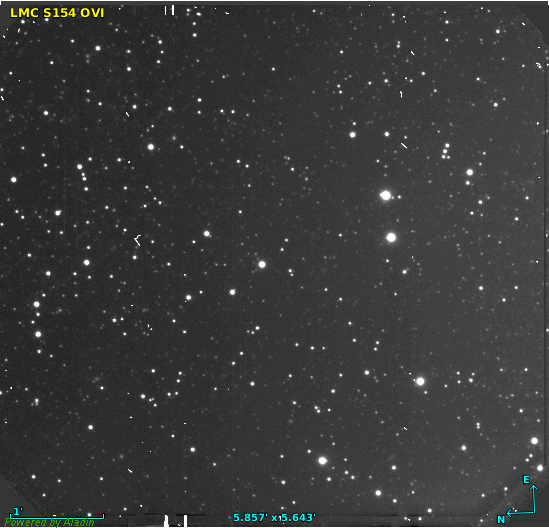}{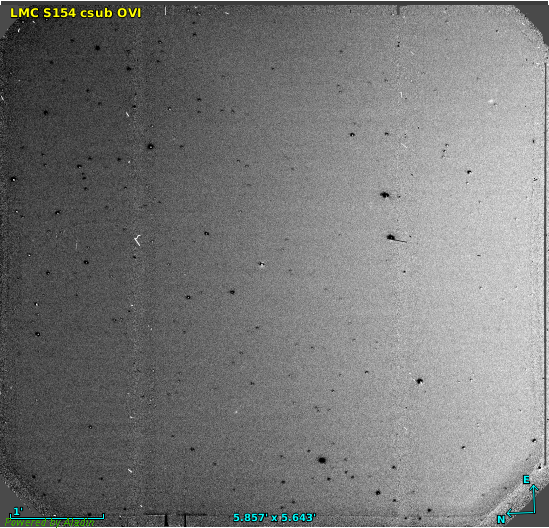}
\caption{As in Fig. \ref{fig:v1016}, but for LH$\alpha$120 S154 (observed on March 15, 2018). Its highly variable Raman OVI emission must have been at its minimum state, or absent, considering our non-detection.\label{fig:lmcs154}}
\end{figure}

Our RAMSES II images taken on March 15, 2018 do not show any significant Raman emission from LH$\alpha$120 S154 (Fig.~\ref{fig:lmcs154}): due to the object strong spectroscopic variability, its Raman OVI 6830 band must have been at its minimum state, or absent, probably following the decreasing trend reported by I{\l}kiewicz et al. (2019).

\subsubsection{V366 Car and GDS J0954243-571655} \label{subsubsec:v366 car}
\begin{figure}[ht!]
\plottwo{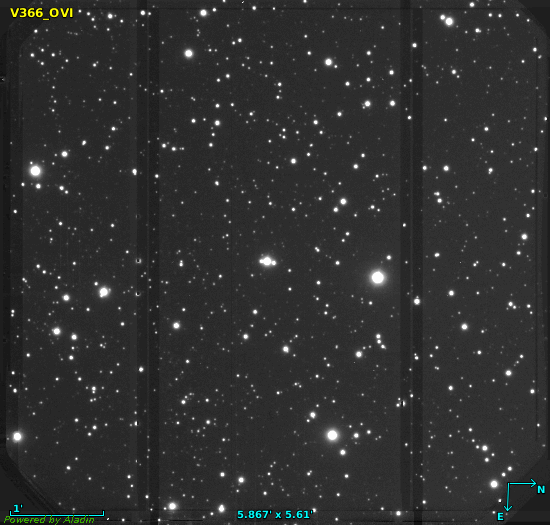}{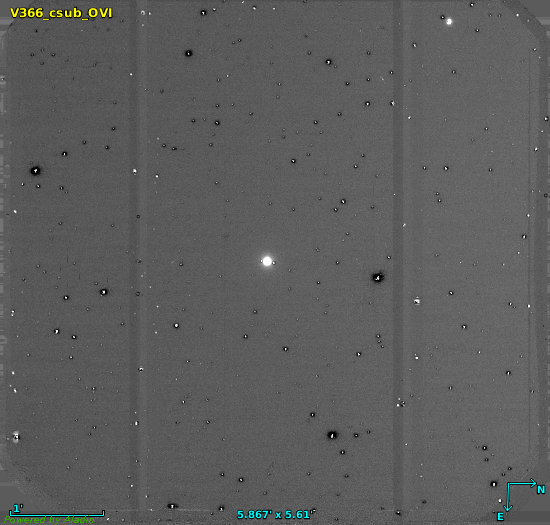}
\caption{As in Fig.~\ref{fig:v1016}, but for V366 Car (observed on March 14, 2018). Also in this case, the target Raman emission was easily recovered. More interesting, however, is the appearance of another clear detection in the continuum-subtracted image, at the top-right corner (see Section~\ref{subsubsec:v366 car} and Fig.~\ref{fig:gdsspec} for details).\label{fig:v366}}
\end{figure}

The last observation we show in this section is that of V366 Car, another fairly well-characterized SySt, whose donor star is an M6 giant (Allen 1984, MZ02).
As we show in Fig.~\ref{fig:v366}, its Raman OVI 6830 emission was promptly recovered by RAMSES II.\\

Noteworthy in this sky-field, however, is that in the OVI continuum-subtracted image another object appears to be clearly detected. A quick check in the public archives allowed us to identify it as GDS J0954243-571655 ($\alpha_{J2000}$=09:54:24.38, $\delta_{J2000}$=-57:16:55.5), an anonymous variable star listed in the Bochum Galactic Disk Survey (Hackstein et al. 2015). The only information we were able to recover about this very red object (2MASS J=7.30, H=6.11, J-K=1.70) is that it is classified by the ASAS-SN survey (Shappee et al. 2014; Kochanek et al. 2017) as a semi-regular variable with an overall amplitude $\Delta$V $\sim$0.6 mag and a tentative period P$\sim$271 days\footnote{The ASAS-SN light curve of GDS J0954243-571655 is publicly available at https://asas-sn.osu.edu/database/light\_curves/393901}.\\

\begin{figure}[hb!]
\plotone{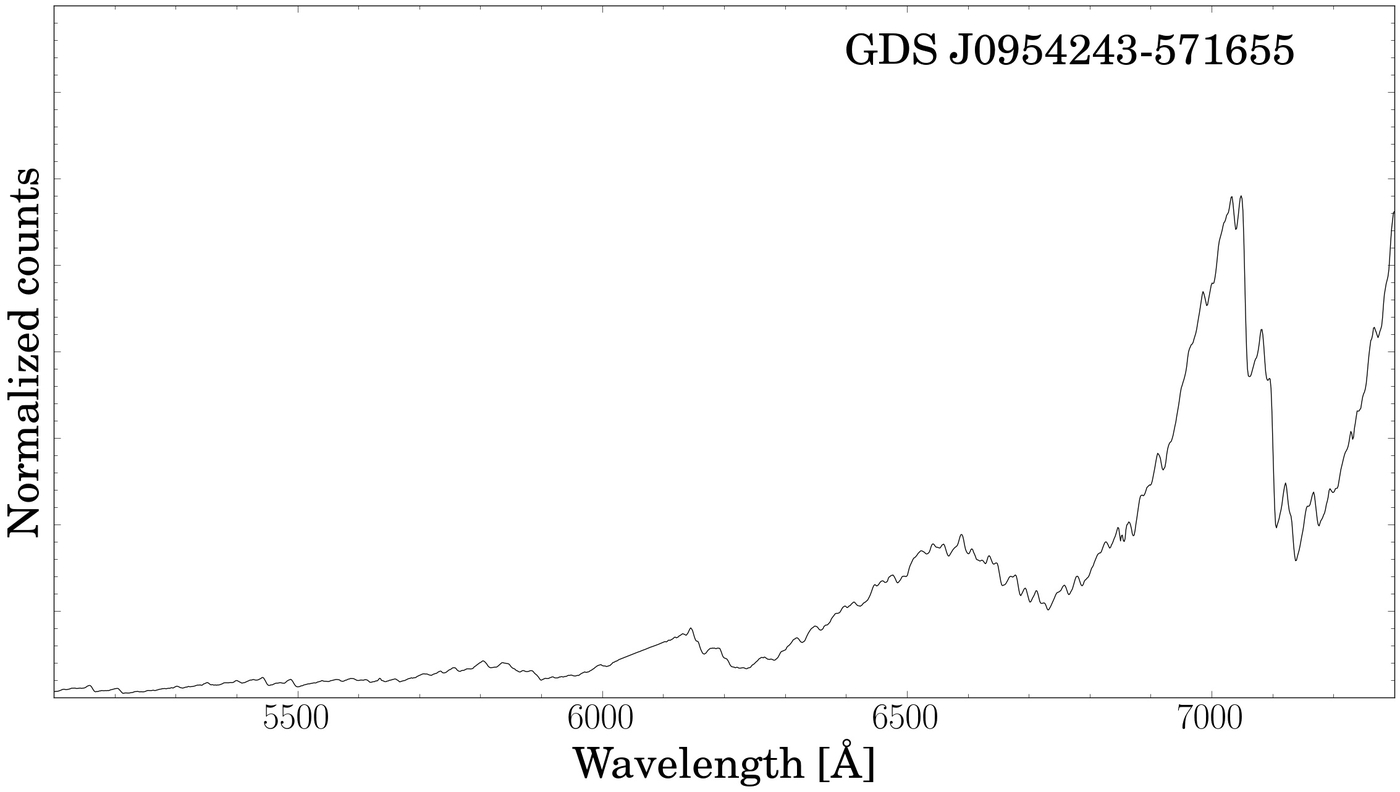}
\caption{GMOS-S spectroscopic follow-up of GDS J0954243-571655, obtained on April 02, 2018. Very late-type oxygen-rich giants like this one might result in spurious Raman OVI (6835-6780) detection. Nonetheless, by combining the RAMSES II filters with additional HeII 4686 and H$\alpha$ 6563 narrow-band imaging, we can easily discard these false positive cases.\label{fig:gdsspec}}
\end{figure}

Because during the same SV phase we could not take further narrow-band images to support the detection, we directly requested a spectroscopic follow-up, executed with GMOS-S on April 02, 2018. The resultant optical spectrum of GDS J0954243-571655 is presented in Fig.~\ref{fig:gdsspec}: quite disappointingly, it shows a very late oxygen-rich giant (as from the relative ratio of the TiO bands and the presence of VO bands) without any symbiotic signature. 
No emission lines are evident: neither H$\alpha$, nor HeII, nor in particular the Raman OVI bands; the positive signal in the subtracted image is in this case due to the steep pseudo-continuum around $\lambda$=6830 \AA\, caused by the particularly strong TiO molecular absorption bands in these very late spectral types. This assessment is empirically confirmed by comparing the ratio between on-band and off-band counts in these RAMSES II images to the ratio between integrated flux around the 6835 \AA\ and 6780 \AA\ regions in the long-slit spectrum (both $\sim$1.3).

In order to avoid this type of issue, when RAMSES II is in full operation, we plan to couple the OVI filters with the HeII and H$\alpha$ ones: with this observing strategy (detailed in Section~\ref{sec:disc}), in the case of GDS J0954243-571655 we would have been immediately able to discard this detection as a spurious one. We postpone to a forthcoming paper a proper characterization of this forgotten variable star (Jaque et al., in preparation).

\subsection{New scientific findings} \label{subsec:newsr}
In this subsection, we present the first (and unexpected) scientific results of the project, and highlight the great potential of RAMSES II for independently discovering and characterizing new SySts. 

\subsubsection{LMC 1} \label{subsubsec:lmc1}
The Raman OVI observations of LMC 1 can be considered the first scientific result of RAMSES II. LMC 1 is a carbon-rich SySt whose rich emission-line spectrum has been said to be reminiscent of RR Tel (Morgan 1992). Interestingly enough, since the time of its discovery more than 25 years ago, it has been known as a SySt without Raman emission: Morgan (1992) noticed that amongst the seven Magellanic SySts for which spectral information in the region near $\lambda$6830 \AA\ was available at that time, LMC 1 was the only object that did not show any Raman features. The spectrum by MZ02 (here reproduced in Fig.~\ref{fig:mzprofile} and taken on October 15, 1994) confirms the absence of the OVI band, as does a more recent X-Shooter/VLT spectrum taken on February 8, 2013 (Soto King, in preparation).
For this reason, we decided to include LMC 1 amongst the targets of our SV phase, and give it high-priority: the idea was to test our method against any false positive, even within the same symbiotic population.\\

\begin{figure}[ht!]
\plottwo{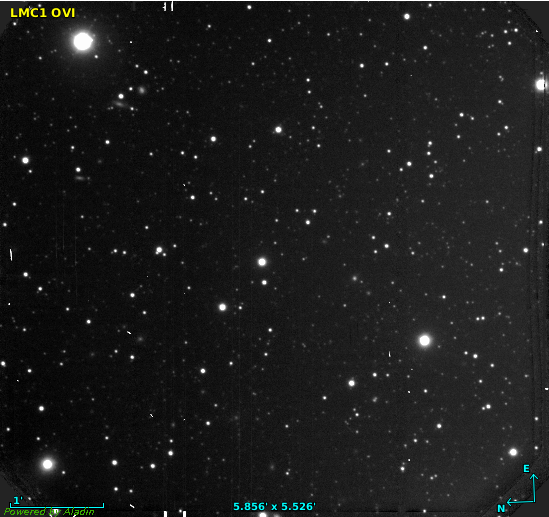}{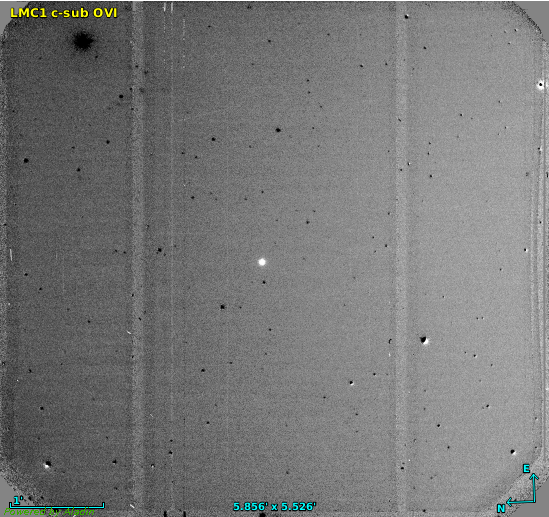}
\caption{As in Fig. \ref{fig:v1016}, but for LMC 1 (observed on March 14, 2018). The Raman detection in this SySt known (and selected in this project) for not showing any Raman emission can be considered the first scientific result of RAMSES II (see Section~\ref{subsubsec:lmc1} for details).\label{fig:lmc1}}
\end{figure}

\begin{figure}[ht!]
\plotone{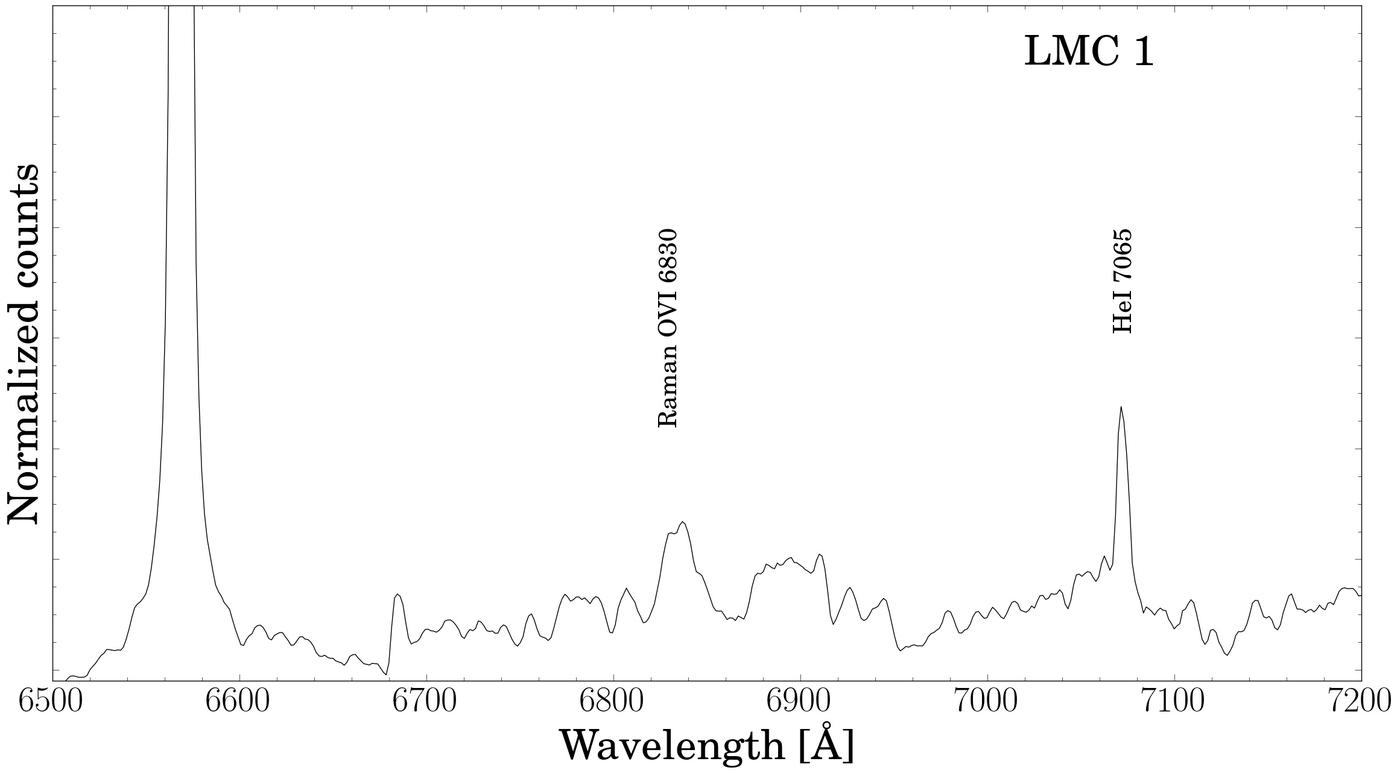}
\caption{GMOS-S follow-up spectrum of LMC 1 obtained on March 15, 2018, i.e., the night following the RAMSES II images shown in Fig.~\ref{fig:lmc1}. The Raman OVI 6830 emission band is evident, as is the change in the intensity of other emission lines when compared with the MZ02 spectrum of Fig.~\ref{fig:mzprofile}.\label{fig:lmc1spec}}
\end{figure}

It was therefore surprising to detect clear Raman emission ($|W_\lambda^{Ram}|\approx$ 13 \AA) during one of the first images of the RAMSES II project, from the very object that had been specifically selected for not displaying Raman emission (Fig.~\ref{fig:lmc1})! 
In order to clarify such a disorienting finding, the following night (i.e., March 15, 2018) we obtained a spectroscopic follow-up with the same GMOS-S, and were thus able to confirm the undisputed appearance of Raman OVI 6830 emission ($\lambda_c$=6835 \AA,  FWHM  $\approx$ 19 \AA, $|W_\lambda|\approx$ 13 \AA) in LMC 1, as shown in Fig.~\ref{fig:lmc1spec}.\\

The two $|W_\lambda|$ values, from spectroscopy and RAMSES II photometry, are in this case virtually identical within their respective uncertainties, demonstrating once more the reliability of our RAMSES II not only in recovering, but also in characterizing the Raman emission in SySts. This spectroscopic confirmation, obtained at the beginning of the SV phase, has immediately corroborated the huge potential of our new method, and has formally opened a new area of study into the temporal behavior of Raman emission in SySts.

\subsubsection{Hen 3-1768} \label{subsubsec:hen3}
Hen 3-1768 (aka ASAS J195948-8252.7) was selected by Lucy et al. (2018) as a symbiotic candidate using photometry from SkyMapper (\textit{uvgriz}), 2MASS, and WISE. In May 2018, we decided to include it at the last minute in the target list of our GMOS-S B4 program.\\

\begin{figure}[hb!]
\plottwo{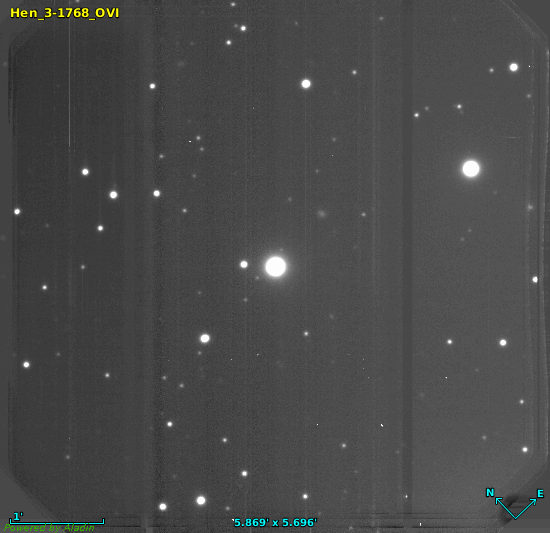}{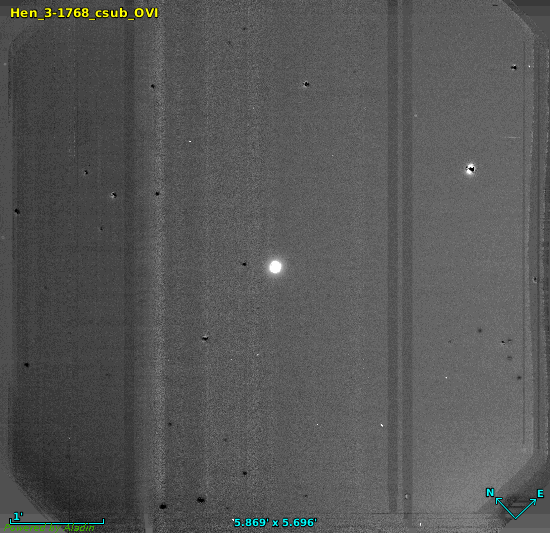}
\caption{As in Fig. \ref{fig:v1016}, but for Hen 3-1768. These images were taken on May 14, 2018, within two days of the first spectrum shown in Lucy et al. (2018), who have spectroscopically ratified the symbiotic nature of this Henize's object.\label{fig:hen3}}
\end{figure}

The RAMSES II images of Hen 3-1768 were taken on May 14, 2018 and are shown here in Fig.~\ref{fig:hen3}. Despite the bad and highly variable seeing during the observation, and some aesthetic issues due to charge smearing effects affecting the instrument detectors on that night\footnote{https://www.gemini.edu/sciops/instruments/gmos/status-and-availability}, the detection of Hen 3-1768 as a Raman emitter ($|W_\lambda^{Ram}|\approx$ 8 \AA) is beyond reasonable doubt.
Considering both its 2MASS infrared colors and its previously reported H$\alpha$ emission (i.e., being listed in the Henize's 1976 survey of southern emission-line stars), the symbiotic nature of the candidate appeared virtually certain.\\

Hen 3-1768 has been very recently confirmed in the traditional way, i.e., using an optical spectrum, to be a yellow SySt by Lucy et al. (2018), who present medium- and low-resolution optical spectra taken in May/June 2018 in which Raman OVI bands ($|W_\lambda|\approx$ 7 \AA), as well as HeII 4686, were particularly evident. Their timely spectroscopic confirmation adds further credibility to RAMSES II's potential of independently discovering and characterizing new SySts.

\subsection{SySt candidates with no Raman OVI emission} \label{subsec:cand}
In this final subsection, we briefly discuss the RAMSES II's view of some independently selected symbiotic candidates: ASASSN-V J081823.00-111138.9, CD-28 10578, NSVS J1444107-074451, and GSC 09276-00130. Like Hen 3-1768, in fact, also these four objects were announced as possible SySt candidates by Lucy and collaborators in the AAVSO Special Notice \#632\footnote{https://www.aavso.org/aavso-alert-notice-632}, and were observed through the Raman OVI and OVIC filters under the B4 program GS-2018A-Q-405.\\

Of these four candidates, three did not show any excess in the continuum-subtracted Raman OVI image; only GSC 09276-00130 did. Since we were at the end of the SV phase, we could not immediately implement the full RAMSES II observing strategy (i.e., coupling the Raman images with the HeII 4686 and H$\alpha$ ones -- see Sect.~\ref{sec:disc}): therefore, we could not exclude a priori the possibility that GSC 09276-00130 represented another false positive case, like GDS J0954243-571655 (Section~\ref{subsubsec:v366 car}).\\

Low-resolution spectra of these four candidates were finally obtained with the Wide Field Reimaging CCD Camera (WFCCD) at the Du Pont telescope,  Las Campanas Observatory, during February 2019. None of these objects turned out to be a Raman OVI emitter (E. Congiu, private communication). 
In particular, GSC 09276-00130 resulted to be an oxygen-rich M giant, and produces a positive detection in the Raman filters just because of the strong TiO photospheric bands that cause the pseudo-continuum in the spectral region of interest to be very steep. It is important to remark that also GSC 09276-00130 (exactly as GDS J0954243-571655) would have been promptly discarded using the complementary HeII and H$\alpha$ images in the planned RAMSES II validation procedure (Section~\ref{sec:disc}).
For the sake of completeness, it is worth mentioning that only the NSVS J1444107-074451 spectrum seems to exhibit very weak Balmer emission (Lucy et al. 2019, in preparation), and remains a possible binary candidate reminiscent somehow of SU Lyn, the prototype of a potential subclass of SySts that are powered purely by accretion (Lopes de Oliveira et al. 2018; Mukai et al. 2016). The other three objects do not show any emission lines in our low-resolution spectra.




\section{Discussion} \label{sec:disc}
In the previous section we showed that our continuum-subtraction imaging technique has proven very robust in characterizing the Raman OVI 6830 emission of a quite heterogeneous sample of galactic and LMC SySts. The RAMSES II optical design (in particular, the virtually identical filter FWHM - Table~\ref{tab:lab}) and the implemented observing strategy (same exposure times for the on-band and off-band frames, taken very close in time) ensure that the PSFs are very similar in most cases (unless, e.g., seeing is highly unstable on time-scale shorter than a few minutes). A simple rescaling factor is therefore sufficient to obtain clear detection in the continuum-subtracted images, without the need for more sophisticated and time consuming difference imaging techniques. Nonetheless, the next step of the project will be to further improve our reduction and analysis techniques by implementing more structured difference imaging algorithms (e.g., optimized PSF matching and skewness transition analysis - Hong et al. 2014).\\

As the case of GDS J0954243-571655 (Section~\ref{subsubsec:v366 car}) and GSC 09276-00130 (Section~\ref{subsec:cand}) suggest, Raman narrow-band imaging alone is not immune from false detection: when the candidate spectral energy distribution is that of an oxygen-rich\footnote{Carbon-rich late giants are of less concern as possible source of false positives because in the spectral region around the Raman OVI 6830 band their continua appear rather flat: see, e.g., Fig. 2 of Matsunaga et al. 2017.}  M giant, the increasingly strong (as a function of spectral sub-type) TiO absorption bands cause the pseudo-continuum around 6800 \AA\ to be particularly steep. The relationship between the $\lambda$=7054 \AA\ TiO band strengths (i.e., M giant subtypes) and RAMSES OVI false positives is not yet well established, and will be the subject of a future, dedicated work. In any case, this potential issue can be easily dispelled by combining the Raman continuum-subtracted image with HeII 4686 and/or H$\alpha$ images. In fact, due to the high ionization potentials involved, all SySts known so far as Raman emitters also show the HeII 4686 \AA\ line in emission (along with H$\alpha$).
This means that in the hypothetical case of a simultaneous detection in the three lines, the contamination by sources that are not \textit{bona fide} SySt goes virtually to zero. In the specific case of the two false positives discussed in this work, neither HeII 4686 nor H$\alpha$ emission is indeed present, thus validating the effectiveness of our combined (Raman OVI + HeII + H$\alpha$), purely photometric, observing strategy.\\

The power and novelty of RAMSES II reside in the ability to promptly identify and independently confirm new SySts. This would be particularly helpful when the number of candidates is high (as is the case for massive photometric surveys -- e.g., Corradi et al. 2008; Rodr{\'{\i}}guez-Flores et al. 2014) and when spectroscopic follow-up of individual sources is too time expensive to be feasible; and/or when the candidate object is so faint as to render any spectroscopic follow-up impossible, even with the largest available facilities.
To give an idea of the exposure times that would be necessary to perform spectroscopic follow-up of SySts in external galaxies, we recall here the case presented by Orio et al (2017). The authors observed CXO J004318.8+412016 in M31 with GMOS-N in long-slit mode (0.75 arcsec slit, B600 grating) for a total of $\sim$4.6 hours. Strong emission lines from the Balmer series, HeI and HeII were detected, but the S/N of the continuum was too low to allow for a precise spectral classification of the V$\simeq$22 magnitude cool component. Clearly, in order to obtain symbiotic spectra in more distant galaxies, exposure times of the order of tens of hours would be necessary \textit{for each and every candidate}, making any systematic population study unrealistic.\\

Of course, RAMSES II is not free from biases. The most obvious (and strongest) one is that not every SySt is a Raman emitter. Growing evidence over the last few years suggests a large hidden population of SySts, i.e., those without shell-burning, and therefore without strong emission lines in the optical spectra (Mukai et al. 2016). 
Nonetheless, as the very first results of RAMSES II presented in this work have highlighted, the current statistics giving the numbers of Raman SySts is in itself strongly biased, mainly by the lack of information on the time variability of the Raman features. Although it is reasonable to assume that RAMSES II is mostly sensitive to shell-burning systems, its intrinsic photometric nature will allow us for the first time to conveniently follow the temporal behavior of Raman emission in SySts, as several cases presented in Section~\ref{sec:sv} have clearly demonstrated.\\

As a matter of fact, Raman OVI features exhibit a clear (ill-studied) temporal variability. Considering the special requirement of their formation in a very thick neutral region in the vicinity of a strong far-UV source, various outburst activities may induce changes in the ionization structure of the binary system leading to variation in the band strength (as in LH$\alpha$120 S154, I{\l}kiewicz et al. 2019). However, no strong dependence of the Raman OVI fluxes on the binary orbital phase has been reported yet, calling for further investigation.

\section{Concluding remarks} \label{sec:remarks}
Based on the analysis of the tests performed during the Acceptance Phase, RAMSES II fully complies with the original requirements set by the IUP: all filters match or exceed the required specifications in terms of cosmetics, physical properties, and optical properties, with total transmission in excess of 90\%.  Continuum-subtracted images using the new Raman OVI filters clearly revealed known SySts with a range of Raman OVI line strengths, even in crowed fields. RAMSES II SV observations also produced the first detection of Raman OVI emission from the SySt LMC 1 and confirmed Hen 3-1768 (selected as a candidate by Lucy et al. 2018) as a new galactic SySt -- the first photometric confirmation of a SySt.
We hope that, as happened with the very same RAMSES II team -- who first gathered during the 2016 Chile-Korea-Gemini Workshop on Stellar Astrophysics\footnote{https://kochil2016.weebly.com/} -- the success of this new methodology will naturally foster national and international collaboration in the field of SySt research.\\

Our team recently imaged three Local Group galaxies in the complementary H$\alpha$ and HeII filters thanks to three observing programs awarded in the 2018B regular CfP (GN-2018B-Q-211, GS-2018B-Q-115 and GS-2018B-Q-219). In addition, we have just completed the 10 hours of telescope time obtained through the IUP (GN-2018B-DD-103 \& GS-2019A-DD-101): this Guaranteed Time was used for imaging in the OVI and OVIC filters the same galaxy fields observed with the 2018B regular programs. The results of this overall $\sim$30 hours of Gemini telescopes time will be presented in a forthcoming series of papers (Gon\c{c}alves et al., in preparation).

In parallel, we plan to keep observing galactic SySt candidates, as they get announced, to better estimate the overall success rate of our method on significantly larger and heterogeneous samples. Larger and larger numbers of candidates are in fact to be expected when data-mining ongoing and future multi-band photometric surveys, like J-PLUS and S-PLUS (Guti\'errez-Soto et al. 2017; Gon\c{c}alves et al. 2015b), SkyMapper (Wolf et al. 2018; Lucy et al. 2018), or LSST, that eventually lie just around the corner. \\

At the time of writing, the RAMSES II filters are ready to be offered to the entire user community of the Gemini Observatory, providing it with the very first tool entirely based on purely photometric criteria for hunting SySts in the local Universe.



\acknowledgments
The authors would like to thank the anonymous referee for helpful comments and suggestions that improved the manuscript. The authors also acknowledge Francisco Campos, who provided a low-resolution spectrum of ASASSN-V J081823.00-111138.9. Based on observations obtained at the Gemini Observatory, which is operated by the Association of Universities for Research in Astronomy, Inc., under a cooperative agreement with the NSF on behalf of the Gemini partnership: the National Science Foundation (United States), National Research Council (Canada), CONICYT (Chile), Ministerio de Ciencia, Tecnolog\'{i}a e Innovaci\'{o}n Productiva (Argentina), Minist\'{e}rio da Ci\^{e}ncia, Tecnologia e Inova\c{c}\~{a}o (Brazil), and Korea Astronomy and Space Science Institute (Republic of Korea).
RAMSES II has been fully funded by a Gemini Observatory Instrument Upgrade Program under the MOU \#20173 (PI: DRG, Observat\'orio do Valongo, Universidade Federal do Rio de Janeiro, Brazil). The RAMSES II team acknowledges the constant support and help of all Gemini staff members involved, at any level, in each phase of this project. 
RA acknowledges financial support from DIDULS Regular PR\#17142 by Universidad de La Serena. DRG acknowledges partial financial support from the CNPq grant 304184/2016-0. SA acknowledges financial support from the Brazilian federal government agency Coordena\c{c}\~{a}o de Aperfei\c{c}oamento de Pessoal de N\'{\i}vel Superior CAPES for a fellowship from the National Postdoctoral Program (PNPD). GJML and NEN are members of the CIC-CONICET (Argentina) and acknowledge support from grant ANPCYT-PICT 0478/14 and 0901/17.  HWL, JEH and BEC are supported by the Korea Astronomy and Space Science Institute under the R\&D program (Project \#2018-1-860-00) supervised by the Ministry of Science, ICT and Future Planning. ABL, JLS, NEN, and GJML gratefully acknowledge Terry Bohlsen's work on spectroscopy of Hen 3-1768. ABL is supported by NSF DGE16-44869 and Chandra DD6-17080X.  JLS is supported by NSF AST-1616646. ABL thanks the LSSTC Data Science Fellowship Program, which is funded by LSSTC, NSF Cybertraining Grant \#1829740, the Brinson Foundation, and the Moore Foundation; their participation in the program has benefited this work. This research has made use of ``Aladin sky atlas'' developed at CDS, Strasbourg Observatory, France. IRAF is distributed by the National Optical Astronomy Observatory, which is operated by the Association of Universities for Research in Astronomy (AURA) under a cooperative agreement with the National Science Foundation. 


\vspace{5mm}
\facilities{Gemini:Gillett (GMOS-N), Gemini:South (GMOS-S), ADS, CDS}
\software{IRAF, Aladin sky atlas, MatplotLib}


\end{document}